\documentclass[twocolumn]{aastex701}
\usepackage{amssymb, amsmath}
\usepackage{natbib}
\usepackage{color}

\definecolor{red}{rgb}{1.0,0.0,0.0}
\newcommand{\matr}[1]{\mathbf{#1}}

\newcommand{\centmass}{1.86 $\pm$ 0.05 $M_{\odot}$}
\newcommand{\centage}{280$^{+140}_{-130}$ Myr}

\begin{document}

\title{SCExAO/CHARIS and Gaia Discovery of a Low-Mass/Candidate Substellar Companion Orbiting The Accelerating Star HIP 46065\footnote{Based in part on data collected at Subaru Telescope, which is operated by the National Astronomical Observatory of Japan.}}


\author[0000-0001-8103-5499]{Taylor L. Tobin}
\affiliation{Lyman Briggs College, Michigan State University, 919 E. Shaw Lane, East Lansing, MI 48825, USA}
\affiliation{Department of Astronomy, University of Michigan, 1085 S. University, Ann Arbor, MI 48109, USA}
\email[show]{tobinta1@msu.edu}

\author[orcid=0009-0001-5554-2435]{Ziying Gu}
\affiliation{Department of Astronomy, School of Science, The Uniersity of Tokyo, 7-3-1 Hongo, Bunkyo-ku, Tokyo 113-0033, Japan}
\email[]{guziying@g.ecc.u-tokyo.ac.jp} 

\author[0000-0002-4677-9182]{Masayuki Kuzuhara}
\affiliation{Astrobiology Center, NINS, 2-21-1 Osawa, Mitaka, Tokyo 181-8588, Japan}
\affiliation{National Astronomical Observatory of Japan, 2-21-1 Osawa, Mitaka, Tokyo 181-8588, Japan}
\email[]{m.kuzuhara@nao.ac.jp}

\author{Dillon Peng}
\affiliation{Department of Physics and Astronomy, University of Notre Dame,
Nieuwland Science Hall, Notre Dame, IN 46556, USA}
\email[]{dpeng@nd.edu}

\author[0000-0002-6845-9702]{Yiting Li}
\affiliation{Department of Astronomy, University of Michigan, 1085 S. University, Ann Arbor, MI 48109, USA}
\email[]{lyiting@umich.edu}

\author[0000-0003-3708-241X]{Aneesh Baburaj} \affiliation{Center for Interdisciplinary Exploration and Research in Astrophysics, Northwestern University, 1800 Sherman Ave, Evanston, IL 60201, USA}
\email[]{ababuraj@northwestern.edu}

\author[0000-0001-5684-4593]{William Thompson}
\affiliation{National Research Council, Herzberg Astronomy and Astrophysics, 5071 West Saanich Road, Victoria, BC, V9E 2E7, Canada}
\email[]{william.thompson@nrc-cnrc.gc.ca}

\author[0000-0001-6305-7272]{Jeffrey Chilcote}
\affiliation{Department of Physics and Astronomy, University of Notre Dame,
Nieuwland Science Hall, Notre Dame, IN 46556, USA}
\email[]{jchilcot@nd.edu}

\author{Mona El Morsy}
\affiliation{Department of Physics and Astronomy, University of Texas at San Antonio, San Antonio, TX 78249, USA}
\email[]{mona.elmorsy@utsa.edu}

\author{Danielle Bovie}
\affiliation{Department of Physics and Astronomy, University of Texas at San Antonio, San Antonio, TX 78249, USA}
\email[]{danielle.bovie@my.utsa.edu}

\author[0000-0002-7405-3119]{Thayne Currie}
\affiliation{Department of Physics and Astronomy, University of Texas at San Antonio, San Antonio, TX 78249, USA}
\affiliation{Subaru Telescope, National Astronomical Observatory of Japan, 
650 North A`oh$\bar{o}$k$\bar{u}$ Place, Hilo, HI  96720, USA}
\email[]{thayne.currie@utsa.edu}

\author[0000-0003-0115-547X]{Qier An}
\affiliation{Department of Physics and Astronomy, Johns Hopkins University, Baltimore, MD 21218, USA}
\email[]{qan4@jh.edu}

\author[0000-0002-1097-9908]{Olivier Guyon}
 \affiliation{Subaru Telescope, National Astronomical Observatory of Japan,
650 North A`oh$\bar{o}$k$\bar{u}$ Place, Hilo, HI  96720, USA}
\affiliation{Astrobiology Center, 2-21-1 Osawa, Mitaka, Tokyo 181-8588, Japan}
\email[]{guyon@naoj.org}

\author{Timothy D Brandt}
\affiliation{Space Telescope Science Institute, Baltimore, MD, USA}
\email[]{tbrandt@stsci.edu }

\author[0000-0002-0457-2941]{Marah Brinjikji}
\affiliation{Department of Physics and Astronomy, University of Notre Dame,
Nieuwland Science Hall, Notre Dame, IN 46556, USA}
\email[]{mbrinjik@nd.edu}

\author[0000-0002-4918-0247]{Robert De Rosa}
\affiliation{European Southern Observatory, Alonso de Córdova 3107, Vitacura, Santiago, Chile}
\email[]{rderosa@eso.org}

\author[0000-0001-5978-3247]{Tyler D Groff}
\affiliation{NASA-Goddard Space Flight Center, Greenbelt, MD USA}
\email[]{tyler.d.groff@nasa.gov}

\author[0000-0001-8345-593X]{Markus Janson}
\affiliation{Department of Astronomy, Stockholm University, AlbaNova University Center, Stockholm, 10691, Sweden}
\email[]{markus.janson@astro.su.se}

\author[0000-0002-6963-7486]{N. J. Kasdin}
\affiliation{Department of Mechanical and Aerospace Engineering, Princeton University, Princeton, New Jersey 08544, USA}
\email[]{jkasdin@princeton.edu}

\author[0000-0002-9420-4455]{Brianna Lacy}
\affiliation{Department of Astronomy and Astrophysics, University of California-Santa Cruz, Santa Cruz, CA, USA}
\affiliation{Oak Ridge Associated Universities, USA}
\affiliation{NASA Ames Research Center, USA}
\email[]{blacy@ucsc.edu}

\author[0000-0003-2426-3852]{Jie Li}
\affiliation{Department of Physics and Astronomy, University of Texas at San Antonio, San Antonio, TX 78249, USA}
\email[]{jie.li@utsa.edu}

\author[0000-0002-3047-1845]{Julien Lozi}
\affiliation{Subaru Telescope, National Astronomical Observatory of Japan,
650 North A`oh$\bar{o}$k$\bar{u}$ Place, Hilo, HI  96720, USA}
\affiliation{Astrobiology Center, 2-21-1 Osawa, Mitaka, Tokyo 181-8588, Japan}
\email[]{lozi@naoj.org}

\author[0000-0002-4164-4182]{Christian Marois}
\affiliation{National Research Council, Herzberg Astronomy and Astrophysics, 5071 West Saanich Road, Victoria, BC, V9E 2E7, Canada}
\email{christian.marois@nrc-cnrc.gc.ca}

\author[0000-0003-4205-4800]{Bertrand Mennesson}
\affiliation{Jet Propulsion Laboratory, California Institute of Technology, Pasadena, California, United States}
\email[]{Bertrand.Mennesson@jpl.nasa.gov}

\author[0000-0001-6177-1333]{Naoshi Murakami}
\affiliation{Astrobiology Center, NINS, 2-21-1 Osawa, Mitaka, Tokyo 181-8588, Japan}
\affiliation{National Astronomical Observatory of Japan, 2-21-1 Osawa, Mitaka, Tokyo 181-8588, Japan}
\affiliation{The Graduate University for Advanced Studies, SOKENDAI, 2-21-1 Osawa, Mitaka, Tokyo 181-8588, Japan}
\affiliation{Faculty of Engineering, Hokkaido University, Kita 13, Nishi 8, Kita-ku, Sapporo, Hokkaido 060-8628, Japan}
\email[]{naoshi.murakami@nao.ac.jp}

\author[0000-0001-6975-9056]{Eric L. Nielsen}
\affiliation{Department of Astronomy, New Mexico State University, P. O. Box 30001, MSC 4500, Las Cruces, NM 88003, USA}
\email[]{nielsen@nmsu.edu}

\author[0000-0002-6650-3829]{Sabina Sagynbayeva}
\affiliation{Department of Physics and Astronomy, Stony Brook University, Stony Brook, NY 11794, USA }
\email[]{sabina.sagynbayeva@stonybrook.edu}

\author[0000-0001-8224-9106]{Garima Singh}
 \affiliation{Subaru Telescope, National Astronomical Observatory of Japan,
650 North A`oh$\bar{o}$k$\bar{u}$ Place, Hilo, HI  96720, USA}
\email[]{gsingh@naoj.org}

\author[0000-0002-9372-5056]{Nour Skaf}
\affiliation{Institute for Astronomy, University of Hawai’i at Manoa, Hilo, HI 96720-2700, US}
\email[]{nskaf@hawaii.edu}

\author[0000-0002-6510-0681]{Motohide Tamura}
\affiliation{Astrobiology Center, 2-21-1 Osawa, Mitaka, Tokyo 181-8588, Japan}
\affiliation{National Astronomical Observatory of Japan, 2-21-1 Osawa, Mitaka, Tokyo 181-8588, Japan}
\affiliation{Department of Astronomy, Faculty of Science, The University of Tokyo, 7-3-1 Hongo, Bunkyo-ku, Tokyo 113-0033, Japan}
\email[]{motohide.tamura@astron.s.u-tokyo.ac.jp}

\author[0000-0002-6879-3030]{Taichi Uyama}
\affiliation{Department of Physics and Astronomy, California State University Northridge, 18111 Nordhoff Street, Northridge, CA 91330, USA}
\email[]{taichi.uyama.astro@gmail.com}

\author[0000-0002-5903-8316]{Alice Zurlo}
\affiliation{Instituto de Estudios Astrof\'isicos, Facultad de Ingenier\'ia y Ciencias, Universidad Diego Portales, Av. Ej\'ercito Libertador 441, Santiago, Chile}
\affiliation{Millennium Nucleus on Young Exoplanets and their Moons (YEMS)}
\email[]{alice.zurlo@mail.udp.cl}

\accepted{2026-11-05}
\submitjournal{\aj}

\shortauthors{Tobin et al.}
\begin{abstract}
We report the direct-imaging discovery of a low-mass (possibly substellar) companion to the astrometrically-accelerating A3V star, HIP~46065 (HD~81254) discovered as part of the Observing Accelerating Stars with SCExAO Imaging Survey (OASIS). SCExAO/CHARIS $JHK$  and Keck/NIRC2 $L_P$ observations of HIP~46065~B data reveal a $JHK$ spectrum consistent with a spectral type of M5$_{-1}^{+3}$: $JHK+L_P$ spectral comparisons to the BT-Settl and SPHINX models support that conclusion, with best fits indicating an object with an effective temperature of $3000^{+100}_{-200}$~K. Orbital modeling of the companion motion across 6 epochs spanning nearly 3 years and host star acceleration with \texttt{orvara} indicate a companion mass of $47^{+75}_{-25}~M_{\rm Jup}$. Finally, we constrain the age of the HIP~46065~A using the upcoming Bayesian inference package for early-type stars (Gu et al., in preparation), deriving a host star mass of \centmass, and an age of \centage.

\end{abstract}

\section{Introduction \label{sec:intro}}

High-contrast imaging provides a unique opportunity to probe the atmospheres and relative orbits of low-mass stellar, substellar, and young planetary-mass companions on Jupiter-like and larger orbits around nearby stars \citep[e.g.][]{Nakajima1995, Chauvin2005a, Marois2008a, Lagrange2010,Currie2023b}. When combined with precision radial velocity and/or astrometry of the host star, the resulting absolute system orbits can be used to constrain the dynamical mass of the companion \citep[e.g.][]{Snellen2018, Brandt2019, Brandt2021b, BrandtM2021, BrandtM2021b}. Such systems, with independent spectral and dynamical mass constraints, can serve as empirical benchmarks for atmospheric and evolutionary models. 

While companions that can be imaged with modern instruments have proven rare \citep{Nielsen2019}, the precision astrometry of Gaia and Hipparcos has opened the door to higher yields of imaged companions through targeted searches.  With measurements spanning $\sim 24$~years, the Hipparcos-Gaia Catalog of Accelerations (HGCA; \citealt{Brandt2021b}) is sensitive to the astrometric acceleration of host stars due to unseen stellar and substellar companions on wide orbits \citep[$P \sim 100$~yrs][]{Brandt2019}. 
Imaging surveys targeting young, nearby ($\lesssim 100$ pc) stars showing astrometric acceleration have resulted in a higher yield of discovered companions \citep{ElMorsy2024}.  This approach significantly bolsters the number of planets \citep[i.e. HIP 99770 b, HIP 54515 b, AF Lep b;][]{Currie2023a,Currie2026,Franson2023} and other substellar companions \citep[e.g. HD 33632 Ab, HIP 71618 B, HD 176585 B, HD 63754 B;][]{Currie2020a,ElMorsy2025,Kuzuhara2022,Li2023,Li2024} to calibrate evolutionary and atmosphere models.

In this paper, we present the discovery of a low-mass companion to the accelerating star, HIP~46065, as part of the Observing Accelerating Stars with SCExAO Imaging Survey (OASIS; PI T.~Currie, Deputy-PI M.~Kuzuhara). In Section \ref{sec:hip46065}, we provide a brief background of the HIP~46065 system. We describe the observations, data reduction, and the retrieval of the companion in Section \ref{sec:obsandredux}. We then analyze the extracted companion spectrum in Section \ref{sec:spec}, comparing it to both empirical libraries (Section \ref{ss:spec-empirical}) and model spectral grids (Section \ref{ss:spec-model}). We present the orbital modeling in Section \ref{sec:orbit}, jointly fitting the relative astrometry of the companion and the host star's absolute acceleration. Finally, in Section \ref{sec:age}, we constrain the age of the HIP~46065 system using the upcoming Bayesian inference package for early-type stars (Gu et al., in preparation).

\begin{deluxetable*}{llllllllll}[ht]
     \tablewidth{0pt}
    \tablecaption{HIP 46065 Observing Log \label{obslog}}
    \tablehead{\colhead{UT Date} & \colhead{Instrument} &  \colhead{Seeing (\arcsec{})} &{Filter} & \colhead{$\lambda$ ($\mu m$)\tablenotemark{a}} 
    & \colhead{$t_{\rm exp}$ (s)} & \colhead{$N_{\rm exp}$} & \colhead{$\Delta$PA ($^{o}$)} & PSF Sub.& SNR}
    \startdata
    2022-02-22 & SCExAO/CHARIS &  0.8 & $JHK$ & 1.16--2.37 & 41.3 & 83 & 19.5 & ADI/ALOCI & 67.3\\
    2023-11-19 & Keck/NIRC2 &  0.6$^{b}$ & $L_{\rm p}$ & 3.78 & 30  & 20  & 5.09 & none & 9.4\\
    2024-03-19 & SCExAO/CHARIS &  --$^{c}$  & $JHK$ & 1.16--2.37 & 31.0 & 29 & 4.8 & SDI/ALOCI & 200.1\\
    2024-05-02 & SCExAO/CHARIS &  --$^{c}$  & $JHK$ & 1.16--2.37 & 31.0  & 29 & 5.0 & SDI/ALOCI & 244.0\\
    2024-11-20 & SCExAO/CHARIS &  0.5 & $JHK$ & 1.16--2.37 &  41.3  &  18 &  2.69 &  none  & 17.2\\
    2024-12-17 & Keck/NIRC2 &  1.6  & $K_{\rm s}$ & 2.16 & 30 & 2  & 0.16 & none & 6.2\\
    \enddata
    \tablenotetext{a}{For CHARIS data, the $\lambda_{\rm \mu m}$ column refers to the wavelength range.  For NIRC2 imaging data, it refers to the central wavelength.}
    \tablenotetext{b}{The AO correction was extremely poor, with Strehl ratios below 0.3.}
    \tablenotetext{c}{No seeing measurements are available but qualitatively the AO correction was representative of slightly below average conditions.}
    \label{tbl:obslog_hip46065}
    \end{deluxetable*}

\begin{figure*}[ht]
    \centering
    \includegraphics[height=5.5cm,trim=  0mm 2.1cm 0mm 0mm,clip]{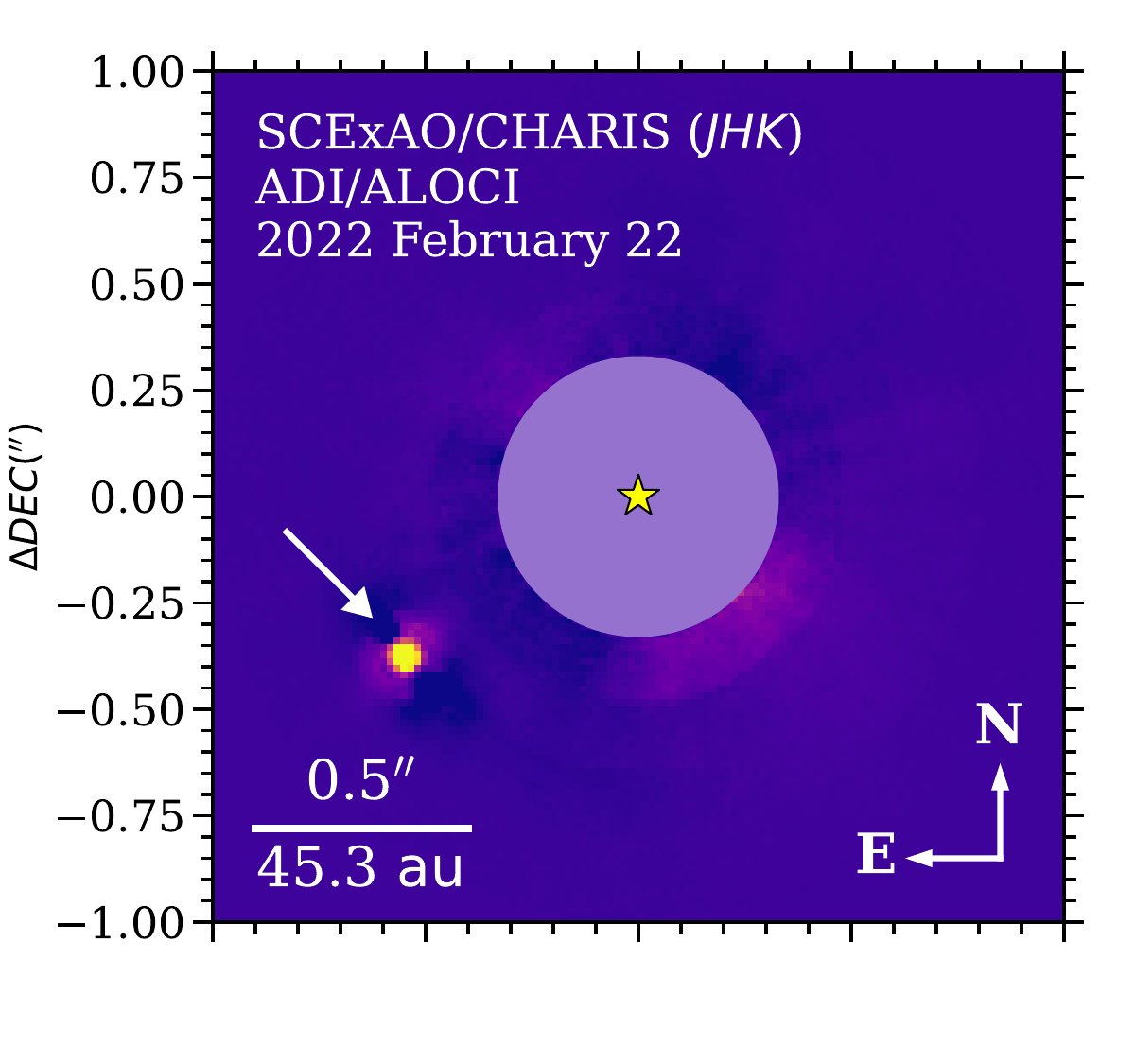}
    \includegraphics[height=5.5cm,trim=3.2cm 2.1cm 0mm 0mm,clip]{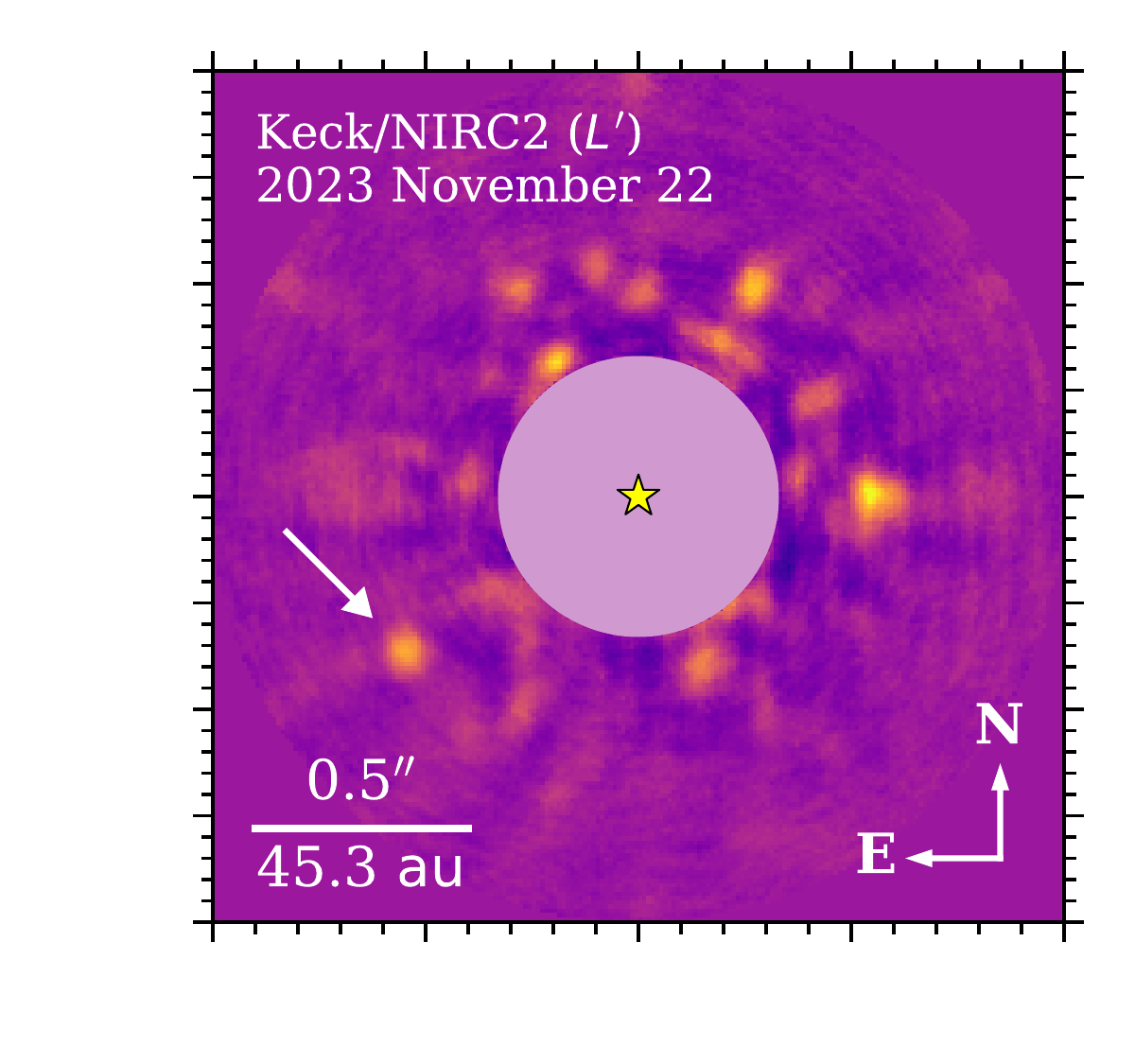}
    \includegraphics[height=5.5cm,trim=3.2cm 2.1cm 0mm 0mm,clip]{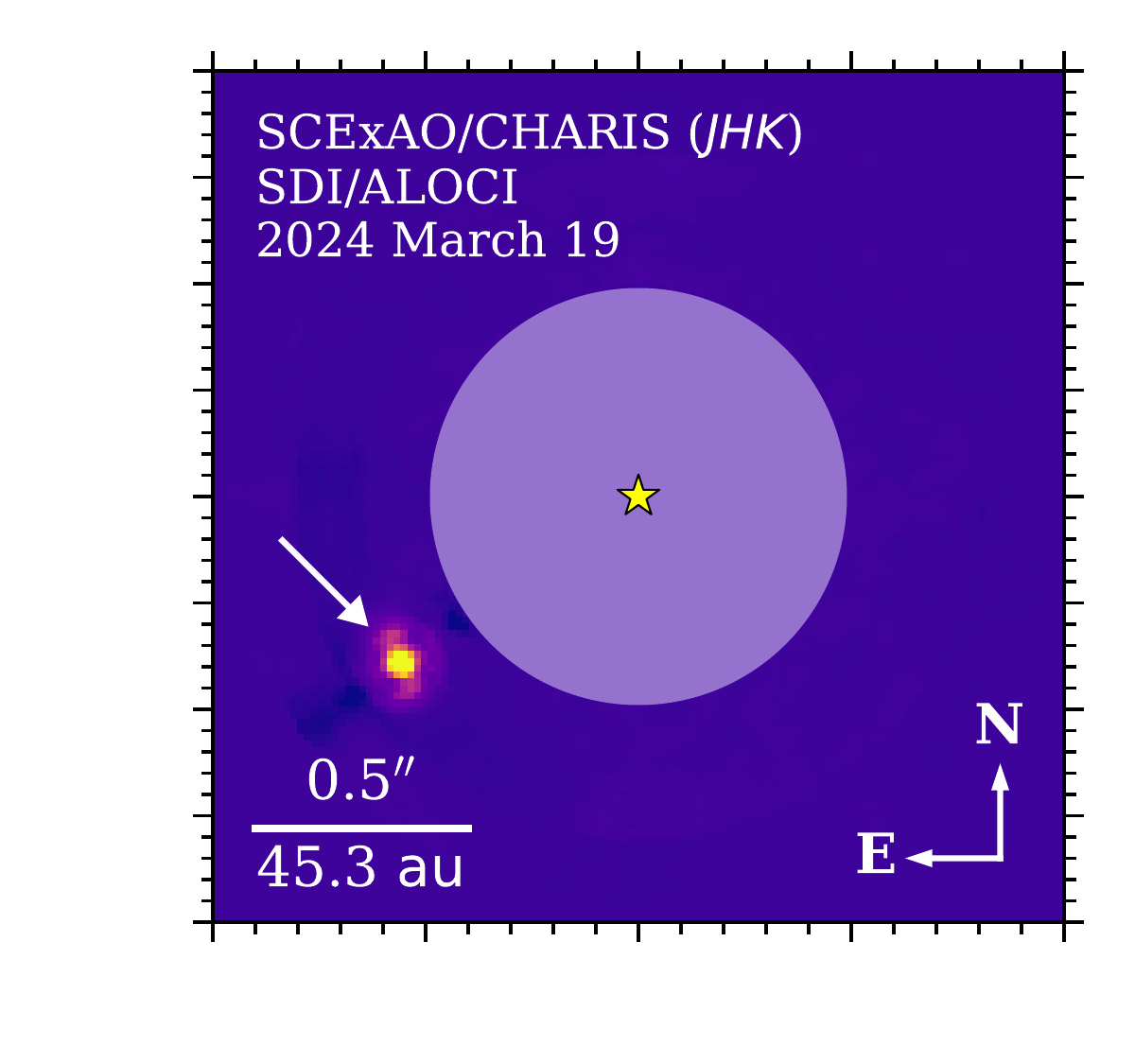}\\
    \includegraphics[height=6.2cm,trim=0mm   0mm 0mm 0mm,clip]{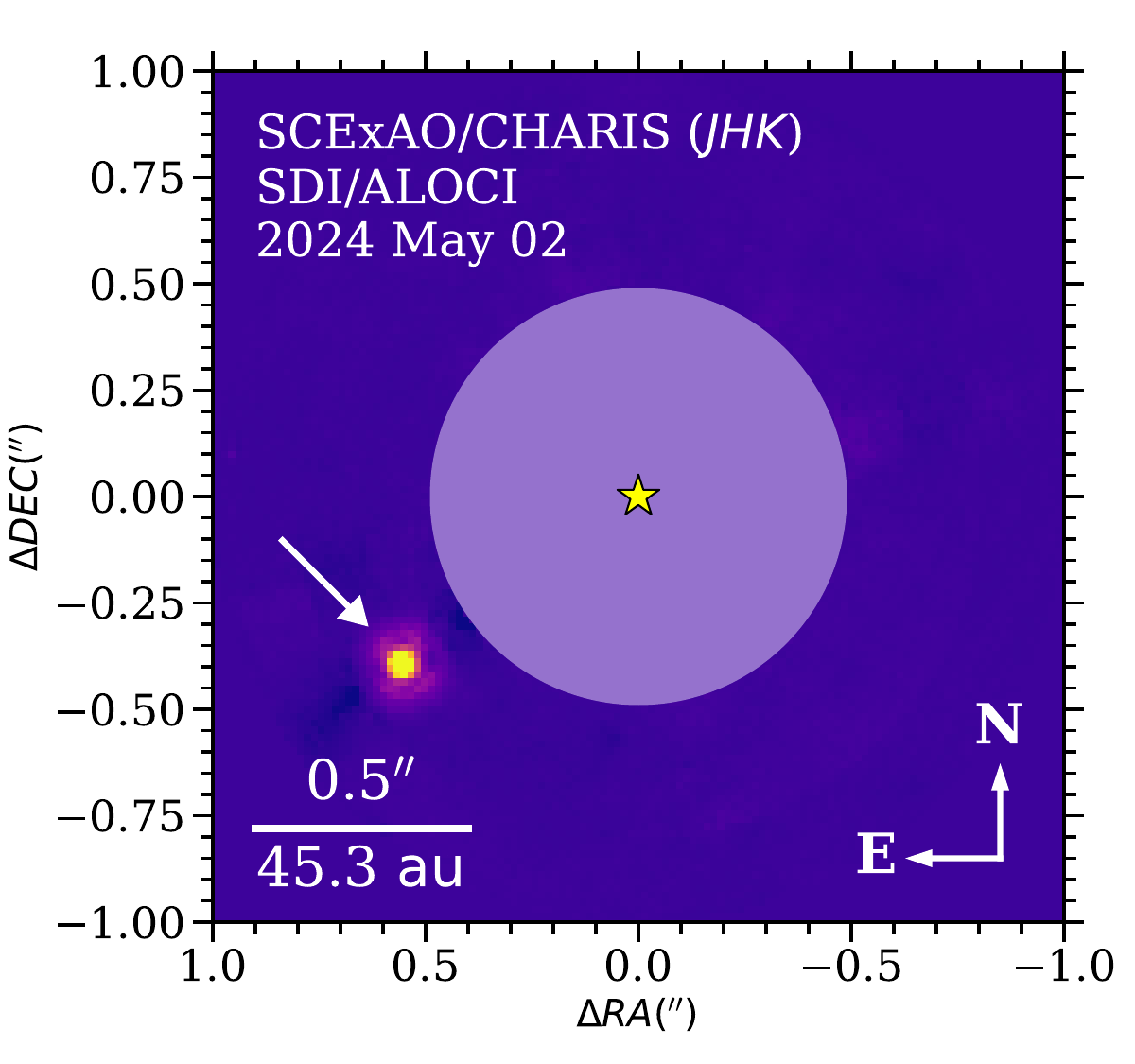}
    \includegraphics[height=6.2cm,trim=3.2cm   0mm 0mm 0mm,clip]{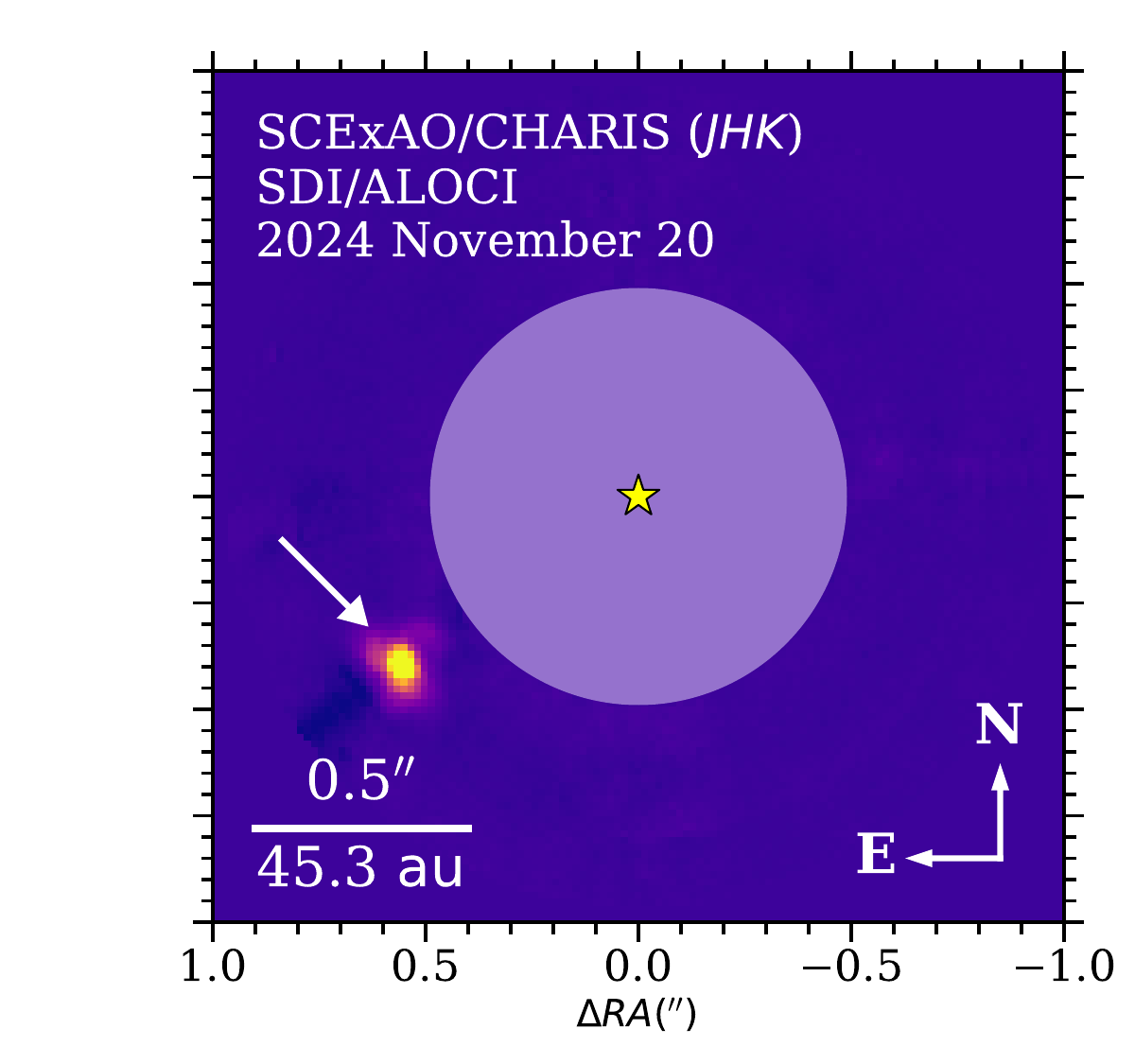}
    \includegraphics[height=6.2cm,trim=3.2cm   0mm 0mm 0mm,clip]{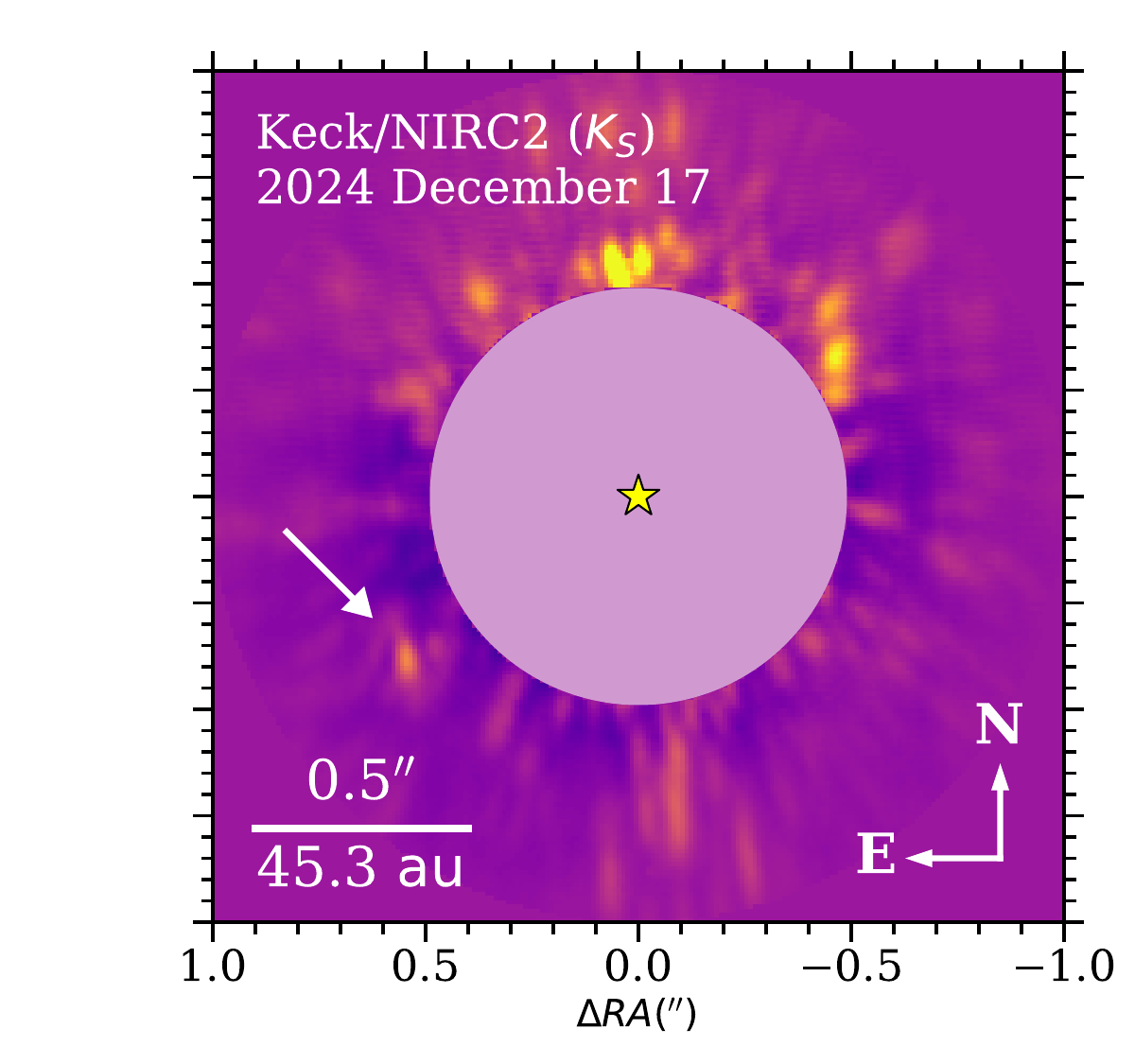}
    
    \caption{Detections of HIP~46065~B in all four SCExAO/CHARIS $JHK$-band epochs and both Keck/NIRC2 epochs listed in Table \ref{tbl:obslog_hip46065}, in order of acquisition. Companion is denoted by white arrow. Position of the central star, HIP~46065~A, is indicated by the yellow star in the center of each frame. PSF subtraction method used for each dataset, if any, is labelled in-figure for each epoch.  For epochs with PSF subtraction applied, the pale circle surrounding the central star indicates the central masked region during the stellar PSF subtraction; for epochs with no PSF subtraction, the masked central region is an arbitrary size for companion visibility. }
    \label{fig:hip46065B_detection}
\end{figure*} 

\section{HIP 46065 System Properties\label{sec:hip46065}}
HIP 46065 (HD 81254) is a bright but relatively unexplored A3V star \citep{Houk1982} located at a distance of $90.50 \pm 0.18$~pc \citep{GAIA2023}.   While Banyan-$\Sigma$ \citep{Gagne2018} does not identify HIP 46065 as a high-confidence member of any known moving group, \citet{Kounkel2019} identify it as a component of the proposed Theia 509 association, which they list as having an age of $\approx$ 200 Myr.  A visual comparison of its Gaia Hertzsprung-Russell Diagram position to Pleiades ($\sim$ 115 Myr old) and Hyades ($\sim$ 700 Myr old) members \citep[e.g. as in][]{Currie2026,ElMorsy2025} places it along the Pleiades locus. An A3V star like HIP~46065 with an age consistent with the Hyades would be approaching the end of its main sequence lifetime. In contrast, an A3V star with the age of the Pleiades would be closer to the zero-age main sequence. Therefore, HIP~46065's consistency with the Pleiades in color-magnitude space places it closer in age to the zero-age main sequence.

The Hipparcos-Gaia Catalog of Accelerations \citep{Brandt2021b} reports that HIP~46065's motion across the sky is inconsistent with that for a single star at the 3.1 $\sigma$ level ($\chi^2 = 12.32$ with 2 degrees of freedom).   The acceleration derived from HGCA is consistent with a companion at 0\farcs{}5 (0\farcs{}25) with a substellar minimum mass of $\approx$ 50 $M_{\rm Jup}$ (7 $M_{\rm Jup}$).  
Thus, we targeted this star as part of the Observing Accelerating Stars with SCExAO Imaging Survey (OASIS; PI T.~Currie, Co-PI M.~Kuzuhara) \citep{ElMorsy2024}.

While this serves as a guidepost to HIP~46065's characterization from previously-published literature, we derive and present additional constraints on HIP~46065~A's metallicity, age, and photometric mass in Section \ref{sec:age}, as well as new constraints on its dynamical mass in Section \ref{sec:orbit}.

\section{Observations and Data Reduction\label{sec:obsandredux}}

\subsection{Observations\label{ss:obs}}
We acquired high-contrast imaging observations of HIP~46065 between February 2022 and December 2024 with the Subaru Telescope -- using the Subaru Coronagraphic Extreme Adaptive Optics (SCExAO) project coupled with the Coronagraphic High Angular Resolution Imaging Spectrograph (CHARIS) integral field spectrograph \citep{Jovanovic2015,Groff2016} -- and with the Keck II Telescope combined with the NIRC2 infrared camera (Table \ref{tbl:obslog_hip46065}). 
The first three SCExAO/CHARIS observations utilized the Subaru facility AO, AO188 \citep{Hayano2008} to provide an initial correction that was then augmented by SCExAO before sharpened starlight was sent to CHARIS.  The November 2024 epoch used the AO3k for this first-order correction.  The first Keck/NIRC2 epoch used the near-IR Pyramid wavefront sensor (WFS) and the second used the facility Shack-Hartmann WFS.  All observations were conducted using pupil-tracking mode to allow for angular differential imaging (ADI; \citealt{Marois2006}).

We operated CHARIS in broadband mode, covering $JHK$ (1.16--2.37 $\mu$m) with a spectral resolution of $R\sim18$ and using a 0\farcs{}113-radius Lyot coronagraph. The first epoch of observations, 22 February 2022, suffered from high-altitude clouds, while the 20 November 2024 data were obtained with splitting of the PSF core and first Airy ring primarily due to low-wind effect \citep{Milli2018} and some AO miscalibration, resulting in lower signal-to-noise than the other epochs. Satellite spots were used for spectrophotometric calibration in all epochs \citep{Jovanovic2015-astrogrids}, with a 25~nm modulation amplitude in the 2024 March epoch and a 50~nm modulation amplitude for the 2022 February and 2024 May and 2024 November epochs. 

The first (November 2023) and second (December 2024) Keck/NIRC2 non-coronagraphic data sets were obtained in the $L_{\rm p}$ and $K_{\rm s}$ filters, using the correlated double sampling mode and multi-correlated double sampling, respectively.  The ($L_{\rm p}$) November 2023 data suffered from a substantially poorer AO correction than typically achieved for bright stars (i.e. Strehl ratios of $\sim$ 0.3--0.35 vs a typical 0.8--0.85 at $L_{\rm p}$) due to AO hardware problems. However, even with the poorer correction, the companion was still easily recoverable by quicklook data processing without the coronagraph; therefore, additional coronagraphic data was not acquired. On the other hand, the December 2024 (NIRC2 $K_{\rm s}$) data consist of only two non-coronagraphic exposures due to reaching azimuth limits for Keck II.

\subsection{Data Reduction\label{ss:data}}

\subsubsection{SCExAO/CHARIS}
CHARIS data cubes were extracted using the CHARIS Data Reduction Pipeline \citep[DRP;][]{Brandt2017}. For the  March and May 2024 epochs of CHARIS data, this was performed by the Automated Data Extraction, Processing, and Tracking System for CHARIS (\texttt{ADEPTS}; \citealt{Tobin2020, Tobin2022}), whereas it was performed manually in other epochs. Subsequent image processing, stellar point-spread function (PSF) subtraction, companion extraction, and forward modelling calibration were performed with the CHARIS Data Processing Pipeline \citep{Currie2020b}. Following sky subtraction and cube registration using the grid of satellite spots, data cubes were spectrophotometrically calibrated using a satellite spot attenuation model and an A3V model atmosphere from the Kurucz library \citep{Castelli2003} as the spectrum of the host star. 

For the 2022 February CHARIS epoch, the stellar PSF was subtracted using ADI/A-LOCI \citep{Currie2012b,Currie2015}. The 2024 CHARIS epochs spanned considerably less rotation ($\sim$0.72 $\lambda$/D--1.33 $\lambda$ at 1.6 $\mu m$ for $\rho$ $\sim$ 0\farcs{}66), which would result in heavy self-subtraction of the companion if processed with ADI. In comparison, CHARIS's broad bandpass means that speckles at $\rho$ $\approx$ 0\farcs{}7 move by up to $\approx$ 9 $\lambda$/D.  Thus, PSF subtraction for those epochs was instead performed using SDI/A-LOCI.  Following \citet{Currie2017}, we performed PSF subtraction in small wedge-like segments.  The reference PSF within a given segment is constructed from an annular region excluding this segment \citep[i.e. \textit{a moving pixel mask}][]{Marois2010b,Currie2012}.  HIP 46065 B is visible in the raw data for all CHARIS data sets, especially in the 2024 epochs.  Thus, we adopted conservative algorithm settings to minimize signal loss: i.e. a rotation gap of 1.5 $\lambda$/D for the February 2022 data and larger gaps for the SDI-processed 2024 data sets ($\delta$ = 2.5--3 $\lambda$/D) \citep[see ][]{Lafreniere2007}.

\subsubsection{Keck/NIRC2}

We used a custom pipeline to reduce the NIRC2 data \citep{Currie2011}.  For the 2023 November $L_{\rm p}$ data, we first performed sky subtraction; the HIP 46065 B PSF core was unsaturated, and thus we registered each image by fitting a gaussian to the star's position and photometrically 
calibrated each image directly using the star's known and measured photometry.   
In the 2024 December data, the PSF core is saturated in all frames.  We performed image registration by cross-correlating a first image with a 180 degree rotation of itself and then cross-correlating the second image with the first. The initial  $180^{\circ}$ cross-correlation procedure exploits the rotational symmetry of the unsaturated wings of the stellar PSF, while allowing for distortions of the PSF along rows or columns due to blooming or charge leakage from the saturated pixels.  Both data sets had poor parallactic angle rotation, but HIP 46065 B was visible in the raw images.  Thus, we simply subtracted a moving-boxed median filter from each image (box width of 25 pixels or $\sim$ 3.2 $\lambda$/D by 3.2 $\lambda$/D), rotated them north-up, and median-combined them to produce a final image.

\subsection{Detection, Photometry, and Spectroscopy of HIP 46065 B\label{ss:detection}}

The companion as seen in the residuals of all wavelength-collapsed CHARIS cubes and the 2023 November NIRC2 data is shown in Fig. \ref{fig:hip46065B_detection}.  In all data sets, the companion signal is not contaminated by residual speckles.  To estimate signal loss due to our PSF subtraction, we used the forward-modeling approach described in \citet{Currie2018}.  We confirmed that HIP 46065 B's processing throughput is near 100\% -- 0.82-0.97 per channel in the February 2022 CHARIS data set and typically 0.96-1.0 in the 2024 datasets -- with negligible astrometric biasing. In other words, there is no significant difference between the companion's actual and apparent position as a result of self-subtraction during PSF processing.

\begin{figure*}[t]
    \centering
    \includegraphics[height=5.5cm]{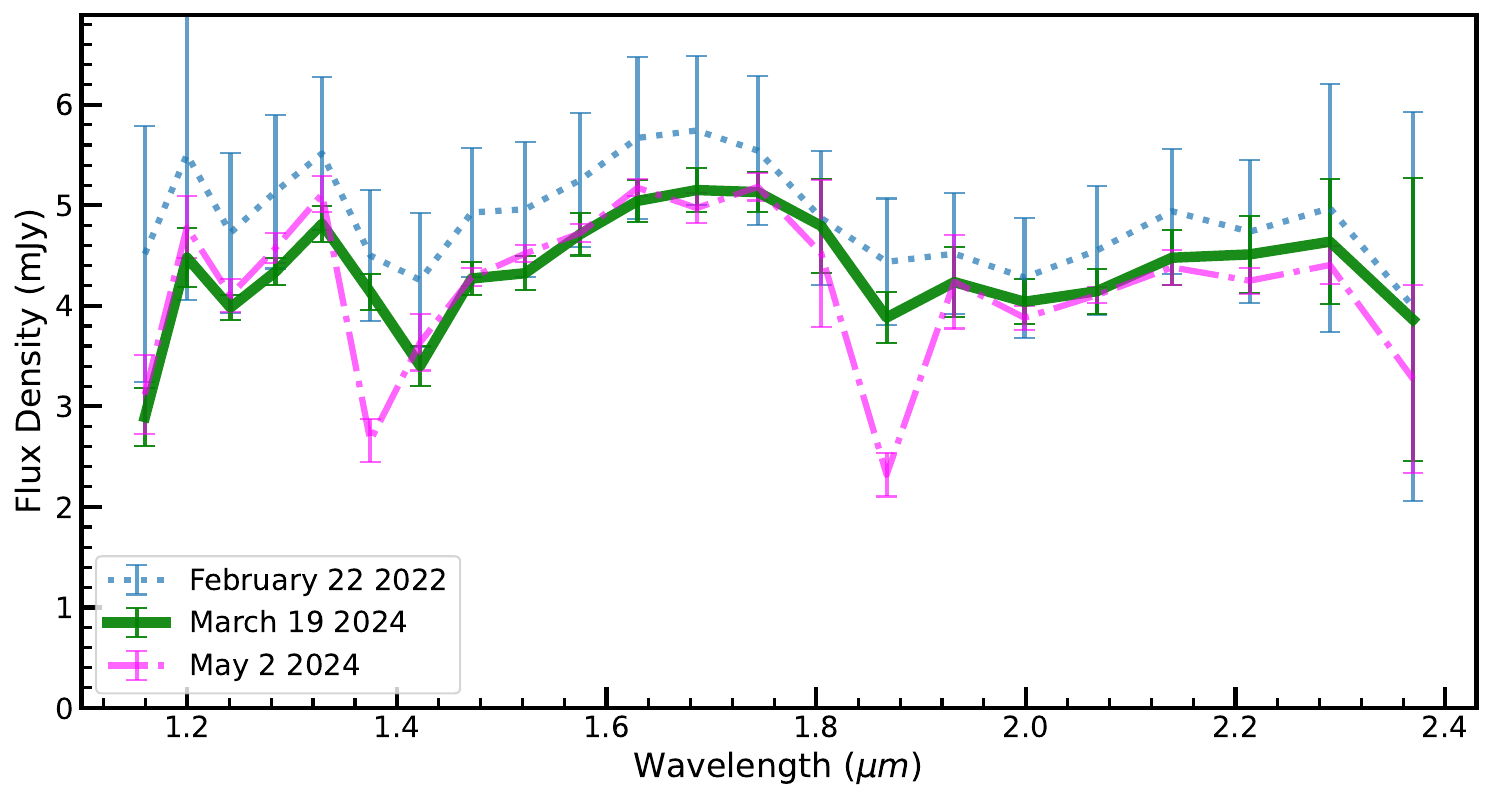}
    \includegraphics[height=5.5cm]{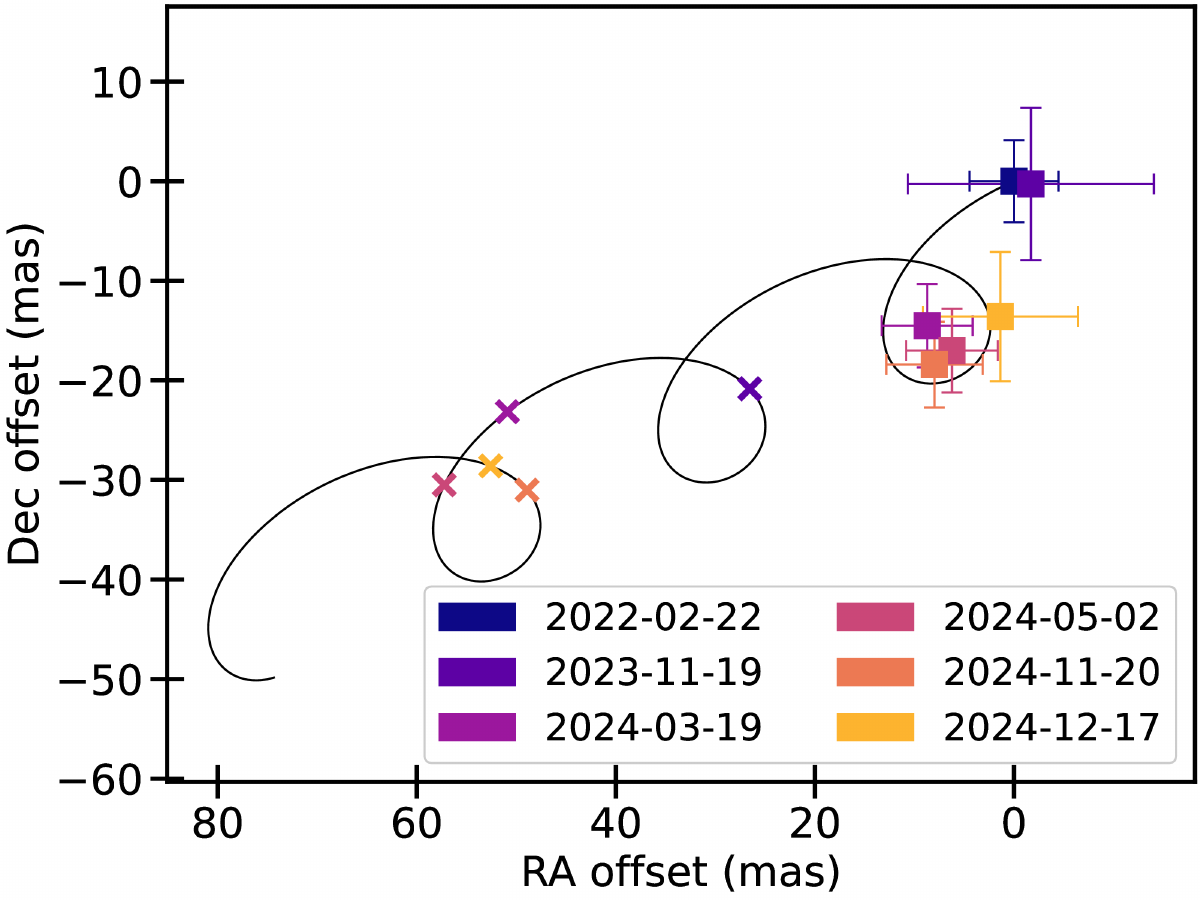}
    \caption{(\textit{Left}) Extracted spectrum of HIP~46065~B in all 3 epochs of SCExAO/CHARIS $JHK$ Broadband observations. (\textit{Right}) The change in measured position of HIP~46065~B across all 6 epochs of observations (colored squares with error bars) compared to that expected for a stationary background source (black curve). Corresponding positions expected for a background source at the measured epoch are given by the `x's of the same color. All positional offsets are with respect to the measured position of HIP~46065~B in the first epoch.\label{fig:spec_pm}}
\end{figure*}

The extracted spectrum from the first three epochs of SCExAO/CHARIS observations is shown in Figure \ref{fig:spec_pm}.  
The extracted companion photometry from the 2023 November NIRC2 epoch, determined from the signal within a PSF-core sized aperture, is $L_P = 12.41 \pm 0.05$~mag.

To establish if the candidate companion is comoving with HIP~46065, we compare the change in position of HIP~46065~B across all 6 epochs with that expected for a stationary background source. We utilize stellar proper motion and parallax values from \citet{GAIADR3}: $\mu_\alpha = -22.600 \pm 0.014$ mas yr$^{-1}$, $\mu_\delta = 9.933 \pm 0.020$ mas yr$^{-1}$, $\pi = 11.0497 \pm 0.0222$ mas. As can be seen in Figure \ref{fig:spec_pm}b, the motion of HIP~46065~B is strongly inconsistent with that of a stationary background source, with multiple measurements $>5-7 \sigma$ away from the expected position of a background object across the nearly 3 years of observations.

\section{Infrared Spectrum and Photometry\label{sec:spec}}

For the SCExAO/CHARIS spectrum, we use the fitting prescription from \citet{GrecoBrandt2016}:

\begin{equation} \label{eq:chisq}
\chi^2 = ( \matr{f_{\nu}} - \alpha_f \matr{F_{\nu}})^T \matr{C}^{-1}( \matr{f_{\nu}} - \alpha_f \matr{F_{\nu}}),
\end{equation}
where $\matr{f_{\nu}}$ is the 1-dimensional matrix of the measured flux density at each wavelength, $\matr{F_{\nu}}$ is same for the model's spectrum, $\alpha_f$ is the dimensionless scaling factor, and $\matr{C}$ is the two-dimensional covariance matrix. The terms of the covariance matrix are parameterized such that:

\begin{align} \label{eq:covelement}
\psi_{ij} \equiv \frac{C_{ij}}{\sqrt{\sigma_i^2 \sigma_j^2}} \approx & A_{\rho} \exp\left[  -\frac{1}{2} \left(\frac{\rho}{\sigma_{\rho}} \frac{\lambda_{i} - \lambda_{j}}{\lambda_{c}}\right)^2 \right] +\nonumber \\
& A_{\lambda} \exp\left[ -\frac{1}{2} \left(\frac{1}{\sigma_{\lambda}} \frac{\lambda_{i} - \lambda_{j}}{\lambda_{c}}\right)^2\right] + A_{\delta} \delta_{ij},
\end{align}
where $\lambda_c$ is the central wavelength for the bandpass (1.65~$\mu$m), $\delta_{ij}$ is the Kronecker delta, and $\rho$ is the companion's separation in units of $\lambda_c/D$. The amplitudes of the correlated and uncorrelated noise terms $(A_{\rho}, A_{\lambda}, A_{\delta})$ and correlation lengths $(\sigma_{\rho}, \sigma_{\lambda})$ are derived with the CHARIS Data Processing Pipeline (DPP; \citealt{Currie2020b}) following \citet{GrecoBrandt2016}. The values derived for these parameters in each epoch are provided in Table \ref{tab:CDPPP}.

\begin{deluxetable*}{lllllll}[t]
 \tablewidth{0pt}
\tablecaption{CHARIS Spectral Covariance Parameters \label{tab:CDPPP}}
\tablehead{
\colhead{UT Date} & 
\colhead{$A_{\rho}$} & 
\colhead{$A_{\lambda}$} & 
\colhead{$A_{\delta}$} & 
\colhead{$\sigma_{\rho}$} & 
\colhead{$\sigma_{\lambda}$} &
\colhead{$\rho$ $(\lambda/D)$}
}
\startdata
20220222 & 0.277 & 0.505 & 0.218 & 0.900 & 0.899 & 15.754\\
20240319 & 0.246 & 0.336 & 0.418 & 0.900 & 0.511 & 15.778\\
20240502 & 0.113 & 0.160 & 0.727 & 0.900 & 1.51 & 15.754\\
\enddata
\tablecomments{Values for the amplitudes of correlated and uncorrelated noise terms ($A$) and correlation lengths ($\sigma$) as calculated by the CHARIS DPP for the observations made using SCExAO/CHARIS.}
\end{deluxetable*}


The spectra from the 2024 SCExAO/CHARIS observations showed significant telluric absorption in several wavelength bins. The affected wavelength bins -- 1.16, 1.37, 1.42, 1.87, and 1.93 $\mu$m (indices 0, 5, 6, 14, and 15) -- are removed when fitting the model and empirical library spectra below.\footnote{The removal of the spectral bins affected by tellurics occurs only during the $\chi^2$ minimization, and is equivalent to down-weighting terms including those wavelengths. The derivation of the covariance matrix following equation \ref{eq:covelement} is still performed using all wavelengths in the CHARIS spectrum.}

\subsection{Comparison to Empirical Libraries\label{ss:spec-empirical}}

We first compare the observed $JHK$ spectrum of HIP~46065B to empirical spectra from the SpeX Prism Spectral Library \citep{Burgasser2014}. Each of the empirical library spectra are first resampled into equispaced wavelength bins with the \texttt{specutils} \texttt{FluxConservingResampler} \citep{specutils, Carnall2017}. We then convolve them with a 1D Gaussian with a standard deviation of $\sigma_{conv} = \left( \sigma^2_{CHARIS}-\sigma^2_{init} \right)^{1/2}$, where $\sigma_{CHARIS}$ is the standard deviation of the CHARIS dispersion and $\sigma_{init}$ is the initial dispersion of the SpeX Prism spectrum. For the CHARIS dispersion, we use the median dispersion in the Broadband filter, $R_{CHARIS}=18.4$, computed at a wavelength of $1.8$~$\mu$m \citep{Groff2016}. Finally, we bin the smoothed model spectra down using the same wavelength bins extracted by the CHARIS DRP \citep{Brandt2017} for the CHARIS Broadband observations.

Following Equation \ref{eq:chisq}, we then determine the scaling factor $\alpha_f$ that minimizes the $\chi^2$ between that empirical spectrum and the observed HIP~46065B spectrum. To account for the uncertainties in the template spectra, the propagated uncertainties of the rebinned empirical spectrum, scaled by the dimensionless scaling factor $\alpha_f$, are added in quadrature to the covariance matrix for each calculation of $\chi^2$ in the minimzation process.

For this analysis, we retrieved all SpeX Prism Spectra with spectral types between K0 and T5 that had a median $SNR \geq 100$ using \texttt{SPLAT} \citep{SPLAT}. Each empirical spectrum is then fit as described above to the HIP~46065B CHARIS $JHK$ spectrum from each epoch. However, the lower $SNR$ of the 2022 epoch provides much weaker constraints on the best-fit spectral type derived through this analysis. Therefore, we focus on the two 2024 epochs.

\begin{figure*}[ht]
    \centering
    \begin{minipage}{.45\linewidth}
            \includegraphics[width=\textwidth]{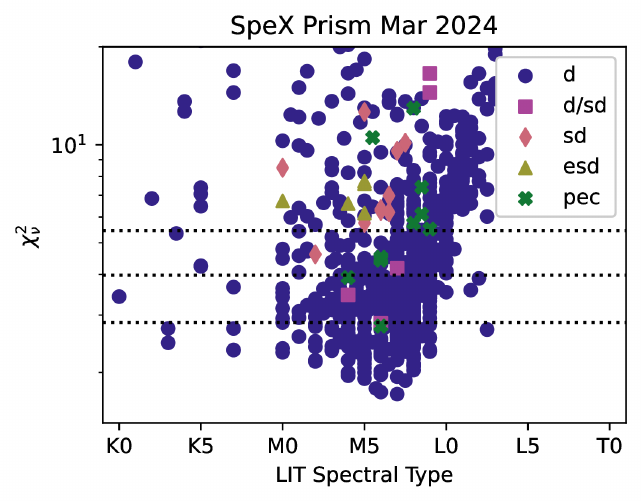}
            \includegraphics[width=\textwidth]{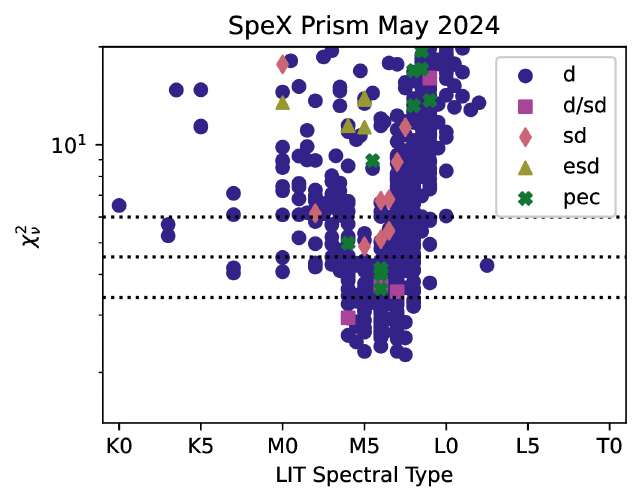}
    \end{minipage}
    \hspace{0.5cm}
    \begin{minipage}{.45\linewidth}
            \vspace{0.6cm}
            \includegraphics[width=\textwidth]{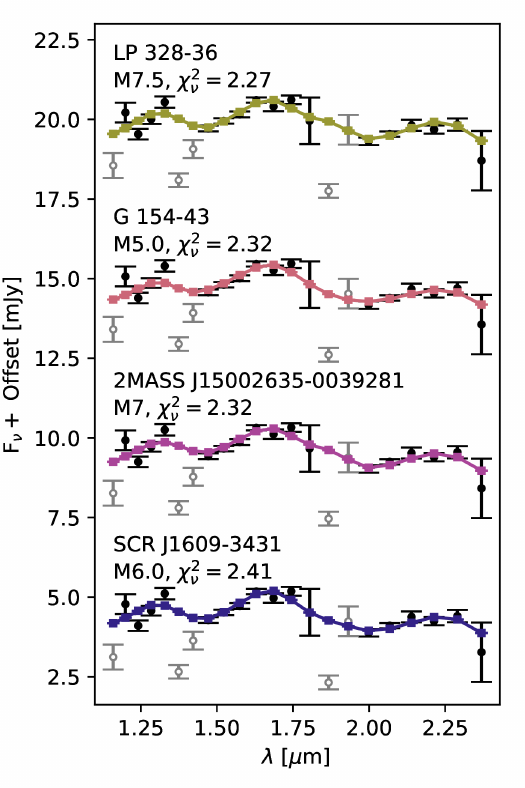}
    \end{minipage}
    
    \caption{Comparison of HIP~46065B CHARIS spectrum to SpeX Prism spectral libraries. (Left) $\chi_{\nu}^2$ for each empirical spectrum as a function of its spectral type resulting from its fit to the March 2024 (upper left) and May 2024 (lower left) HIP 46065~B epochs. Marker color and shape corresponds to indicated spectral classes: dwarf (``d''), dwarf/subdwarf (``d/sd''), subdwarf (``sd''), extreme subdwarf (``esd''), and peculiar (``pec''). Black dotted lines indicate the $\left\lbrace 1, 3, 5 \right\rbrace \times \sigma$ cutoffs relative to the best-fit $\chi_{\nu}^2$. (Right) The four SpeX Prism library spectra yielding the best fit to the May 2024 HIP~46065~B spectrum. Black points with error bars show the measured HIP~46065~B spectral points that were used for the fit, while the gray open circles indicate the points that were masked from fitting due to telluric contamination. Colored points indicate the empirical library spectrum. }
    \label{fig:empspec_chi2_mar2024}
\end{figure*}

The $\chi^2_{\nu}$ fits as a function of spectral type for both 2024 CHARIS epochs are shown in Figure \ref{fig:empspec_chi2_mar2024}, as well as the four best-fit empirical spectra for the March 2024 HIP 46065~B epoch. To compute the $\chi^2_{\nu}$ values that correspond to the family of solutions that fall within $\left\lbrace 1, 3, 5 \right\rbrace \times \sigma$ of the best-fit spectrum for each epoch, we use the ``projection'' method of \citet{Avni1976}. We use the survival function to convert a position some number of standard distributions from the peak of a Gaussian distribution to a $\Delta \chi^2 = \chi^2 - \chi^2_{min}$ within a $\chi^2$ distribution centered on $\chi^2_{min}$ for that epoch of observations. As this determines a family of consistent solutions rather than matching a model with unique solutions for each combination of parameters, we use the same number of degrees of freedom (DoF) as was used for the individual $\chi^2$ fits, 16. The $\left\lbrace 1, 3, 5 \right\rbrace \times \sigma$, or rather $\left\lbrace 68\%, 99.7\%, 99.99994\%\right\rbrace$ confidence intervals, then correspond to $\Delta \chi^2 = \left\lbrace 18.11, 36.22, 59.76 \right\rbrace$ with 16 degrees of freedom, or for comparison with the $\chi^2_{\nu}$ values in Figure \ref{fig:empspec_chi2_mar2024},
$\Delta \chi^2_{\nu} = \left\lbrace 1.13, 2.26, 3.74 \right\rbrace$. 

The library spectrum that best matches the HIP~46065~B spectrum from the March 2024 epoch is the SDSS J153619.14+330515.1 \citep{West2008} with $\chi_{\nu}^2 = 1.72$. However, the empirical sources within the $68\%$ confidence interval ($1\sigma$) cover a broad range, spanning from spectral types of K3 to L2.5.

The May 2024 spectrum yields a similar best-fit spectral type, achieving a minimum $\chi_{\nu}^2 = 2.27$ with the M7.5 dwarf LP~328-36 \citep{BardalezGagliuffi2019}. However, the fit to this epoch yields stronger constraints on the spectral type, with the only sources achieving within the $68\%$ confidence interval ($1\sigma$) of the best-fit $\chi^2_{\nu}$ having spectral types from M4-M8.

The SpeX Prism library also includes designations of some sources as subdwarfs, extreme subdwarfs, and peculiar spectral types. Comparing the empirical fits across these sub-types (e.g. Fig. \ref{fig:empspec_chi2_mar2024}), we find that our HIP~46065~B spectra show a poor fit to those labelled as peculiar at a $>1\sigma$ level, and an even poorer fit to known subdwarfs ($>3 \sigma$) and extreme subdwarfs ($>5 \sigma$).

\subsection{Comparison to Model Spectra\label{ss:spec-model}}

\begin{deluxetable*}{lcccccccc}[t]
 \tablewidth{0pt}
\tablecaption{Model Best Fits \label{tab:model_fits}}
\tablehead{
\colhead{Model Grid} & 
\colhead{$T_{eff}$ (K)} & 
\colhead{log(g)} & 
\colhead{[M/H]} & 
\colhead{C/O\tablenotemark{a}} & 
\colhead{$\alpha$\tablenotemark{b}} &
\colhead{$\chi^2_{\nu}$} &
\colhead{R $(R_{J})$} &
}
\startdata
BT-Settl (Feb 22) & 3000 & 4.5 & 0.0 & N/A & 0.0 & 0.2791 & 2.120\\
SPHINX (Feb 22) & 3000 & 5.5 & 1.0 & 0.7 & N/A & 0.3044 & 2.053\\
\tableline
BT-Settl (Mar 24) & 2900 & 4.0 & -0.5 & N/A & 0.0 & 2.8937 & 2.109\\
SPHINX (Mar 24) & 2900 & 5.5 & 0.5 & 0.7 & N/A & 2.6674 & 2.077\\
\tableline
BT-Settl (May 24) & 2900 & 4.0 & -0.5 & N/A & 0.0 & 3.0366 & 2.123\\
SPHINX (May 24) & 3000 & 5.5 & 1.0 & 0.7 & N/A & 3.9454 & 1.921\\
\enddata
\tablecomments{Values for the best model fits for each epoch of observation. Provided $\chi^2_{\nu}$ values reflect the fit of the individual model spectrum specified to the observation, and are therefore reduced using the original $\textrm{DoF}=17$ for the individual model fits. The values of [M/H], C/O, and $\alpha$ are provided only for identification of the best-fit spectrum, and not to indicate constraints on their actual values.}
\tablenotetext{a}{Variation of C/O ratio only used by SPHINX model grid, not BT-Settl (AGSS2009).}
\tablenotetext{b}{Variation of $\alpha$ enhancement value only used by BT-Settl (AGSS2009), not SPHINX.}
\end{deluxetable*}

\begin{figure*}[ht]
    \centering
    \includegraphics[width=0.48\textwidth,trim={0.7cm 0 1.0cm 0},clip]{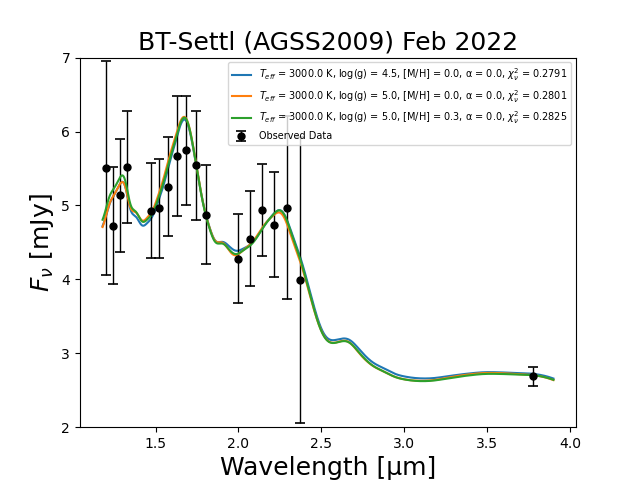}
    \includegraphics[width=0.48\textwidth,trim={0.7cm 0 1.0cm 0},clip]{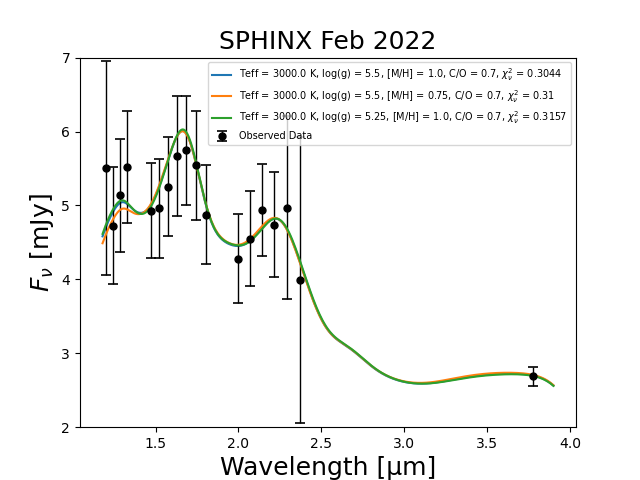}
    \includegraphics[width=0.48\textwidth,trim={0.7cm 0 1.0cm 0},clip]{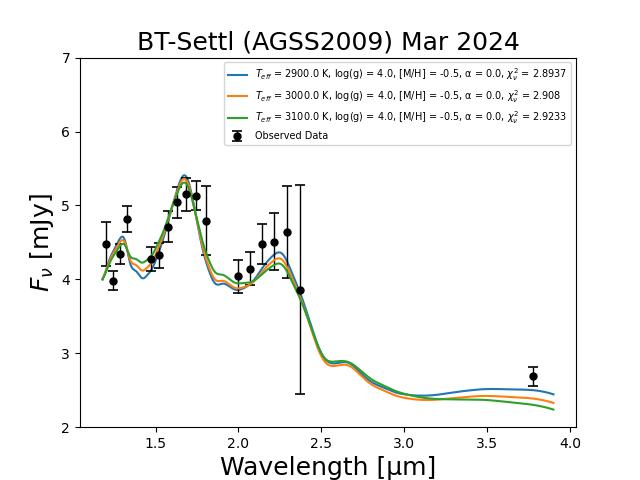}
    \includegraphics[width=0.48\textwidth,trim={0.7cm 0 1.0cm 0},clip]{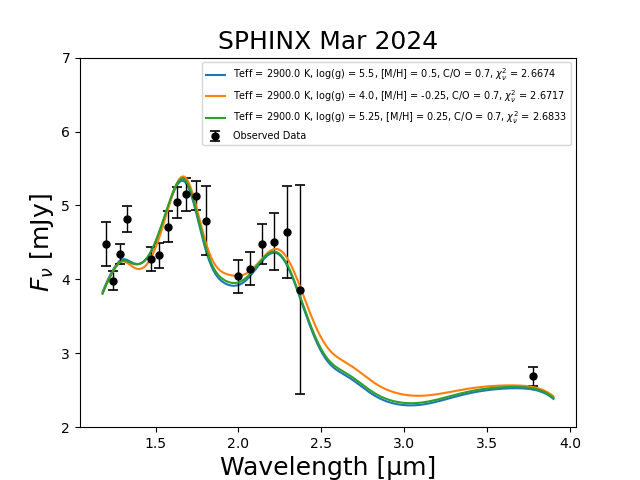}
    \includegraphics[width=0.48\textwidth,trim={0.7cm 0 1.0cm 0},clip]{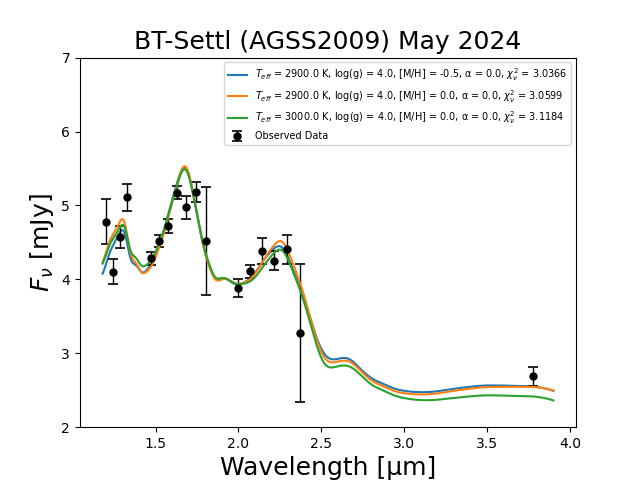}
    \includegraphics[width=0.48\textwidth,trim={0.7cm 0 1.0cm 0},clip]{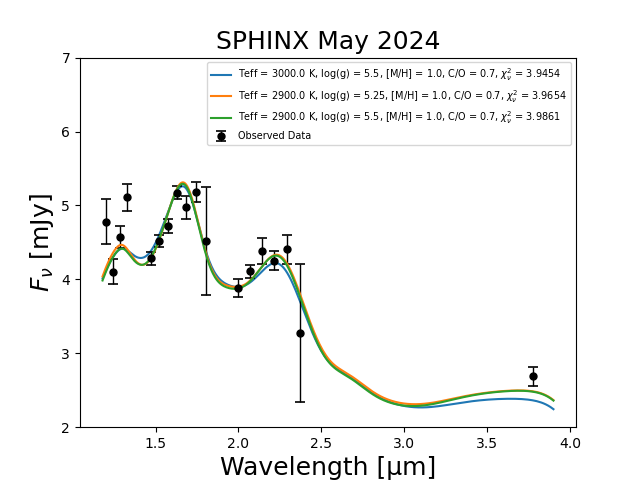}
    
    \caption{Plots of the three best-fit models for BT-Settl (AGSS2009) (left) and SPHINX (right) across all 3 epochs of observation, along with their reduced chi-squared values.}
    \label{fig:Best_Model_Fits}
\end{figure*}

We then compare the measured spectrum of HIP~46065~B to several model atmosphere grids. This fit is performed jointly to the CHARIS $JHK$ spectrum and the L$_p$ photometry point from NIRC2 following the procedure from \citet{Tobin2024}. The model spectra are re-sampled, convolved to the median CHARIS Broadband dispersion, and binned to the resolution of our CHARIS Broadband observations as described in Section \ref{ss:spec-empirical}. The corresponding L$_p$ photometry point is integrated over the filter's bandwidth, using the Keck/NIRC2 L$_p$ transmission curve.\footnote{Retrieved from SVO2 Filter Profile Service \citep{Rodrigo2012, Rodrigo2020, Rodrigo2024}, as sourced from the Keck/NIRC2 instrument documentation.} The uncertainty in the single L$_p$ photometry point does not have any covariance with the independent spectral measurements from CHARIS. Therefore, it can be viewed as an additional term of $(f_{\nu,L_p} - \alpha_f F_{\nu,L_p})^2 / \sigma^2_{F_{\nu,L_p}}$ added to the $\chi^2$ calculation in Eq. \ref{eq:chisq}. 

In contrast to the empirical spectral comparison from Section \ref{ss:spec-empirical}, the model spectra do not have associated spectral uncertainties. Therefore, the uncertainty terms used in this fit -- $\matr{C}$ and $\sigma_{F_{\nu,L_p}}$ -- contain only the uncertainties from the HIP~46065~B measurements.

The models discussed in this paper are the BT-Settl \citep{2011ASPC..448...91A} atmospheric model using the AGSS2009 solar abundances \citep{Asplund2009}, and the SPHINX M-dwarf spectral grid \citep{Iyer_2023, iyer_2023_11392341}. While several other variations of BT-Settl with different solar abundance values were also fitted during the analysis, the difference in best fits was minimal, so we choose to focus on the AGSS2009 models.

BT-Settl is a set of parameter-free atmospheric models based on the PHOENIX code. The BT family of models are generally upgrades over previous models like AMES, with the BT nomenclature representing the incorporation of the updated BT2 water line list \citep{Barber2006}. BT-Settl itself is an improvement on predecessor BT-NextGen with updates to various opacity line lists, solar abundances, and dust and cloud modeling \citep{Allard2012}. Developed specifically for objects ranging from solar down to planetary mass, the BT-SETTL models cover a range of $400 K \leq T_{eff} \leq 70,000 K $ in temperature, $-0.5 \leq \log(g) \leq 5.5 $, and $-4.0 \leq [M/H] \leq 0.5 $ for metallicity. Notably, \citet{Allard2012} mentions that the revision of solar abundances and cloud models in BT-Settl helped fix the tendency that BT-NextGen models had of overestimating effective temperatures of objects below 3500 K, which is a relevant issue when working with objects in the M-L dwarf regime. Note that for some of the models in the BT-Settl library, an $\alpha$-element enhancement value ranging from $0.0 - 0.4$ is an included parameter, which refers to the presence of alpha elements like C, O, Mg, Si, etc. This parameter is not to be confused with the scaling $\alpha_f$ value associated with the reduced chi-squared fitting.

SPHINX is a low-resolution (R $\sim$250), self-consistent 1D atmospheric grid created specifically for modeling M–dwarfs. SPHINX is built upon ScCHIMERA \citep{Line_2013, Arcangeli_2018}, an exoplanet atmospheric retrieval tool. SPHINX uses the EXOPLINES database of precomputed molecular absorption lines, which is more up-to-date compared to the other models referenced in this paper and calculated specifically for objects in the M-dwarf regime \citep{Gharib-Nezhad_2021}. Due to its specialization in M-dwarfs, SPHINX covers a very narrow range of parameters compared to the BT-Settl models, spanning from $2000 K \leq T_{eff} \leq 4000 K$ in temperature, $4.0 \leq$ log(g) $\leq 5.25$, $-1.0 \leq [M/H] \leq 1.0$ in metallicity, and $0.3 \leq$ C/O $\leq 0.9$ in carbon-oxygen ratio. Dust opacities are included in the model, but the atmospheres are assumed to be cloud-free in the version used for analysis in this paper (version 4, last updated in February 2023).  While relatively new, SPHINX has been benchmarked and found to be in agreement with widely used models such as PHOENIX-ACES and BT-Settl on a variety of objects \citep{Iyer_2023}.

\begin{figure*}[h!]
    \centering
    \includegraphics[width=0.47\textwidth,trim={0.7cm 0.3cm 1.4cm 0.6cm},clip]{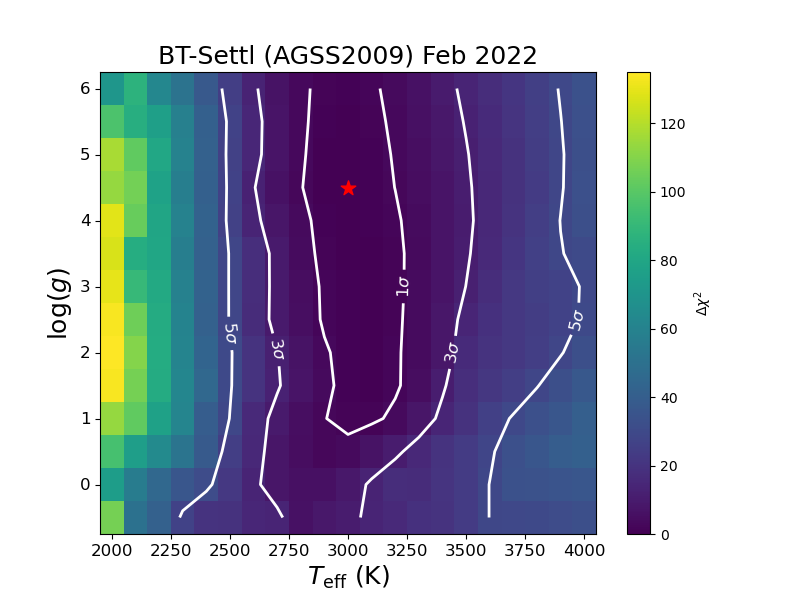}
    \includegraphics[width=0.47\textwidth,trim={0.7cm 0.3cm 1.4cm 0.6cm},clip]
    {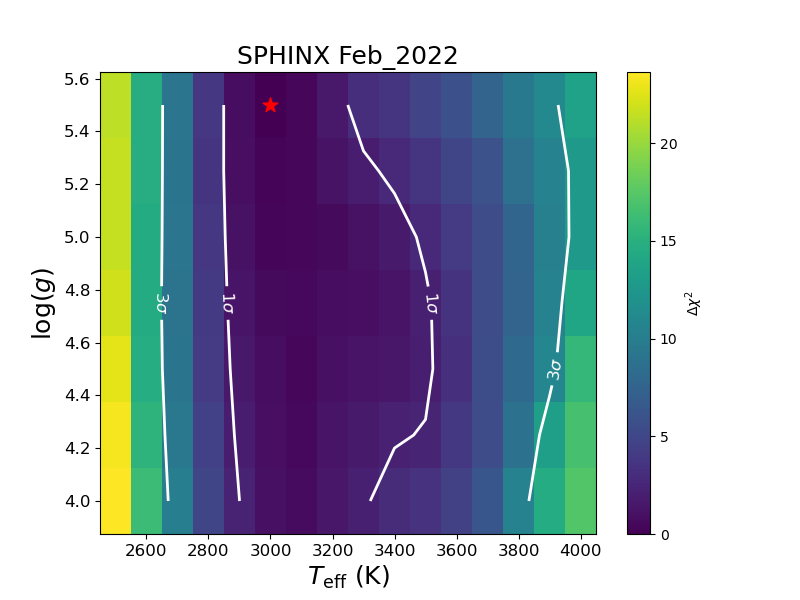}
    \includegraphics[width=0.47\textwidth,trim={0.7cm 0.3cm 1.4cm 0.6cm},clip]
    {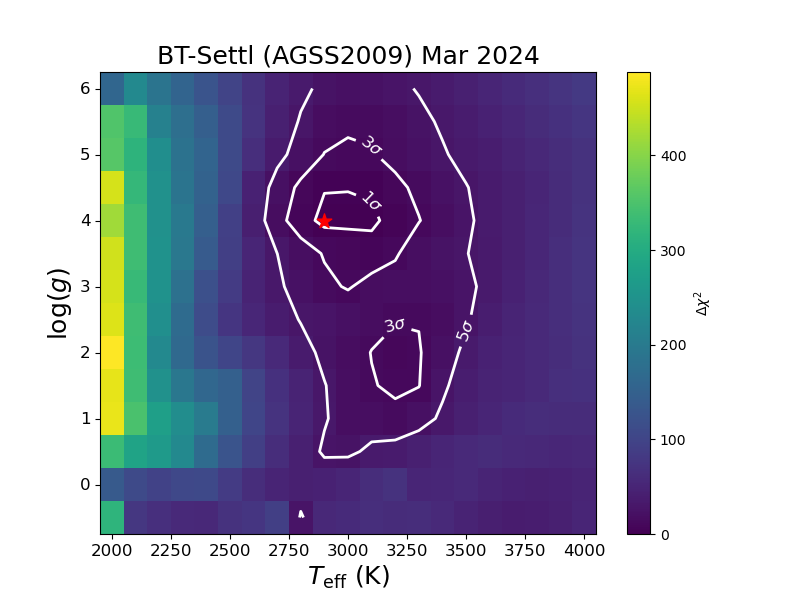}
    \includegraphics[width=0.47\textwidth,trim={0.7cm 0.3cm 1.4cm 0.6cm},clip]
    {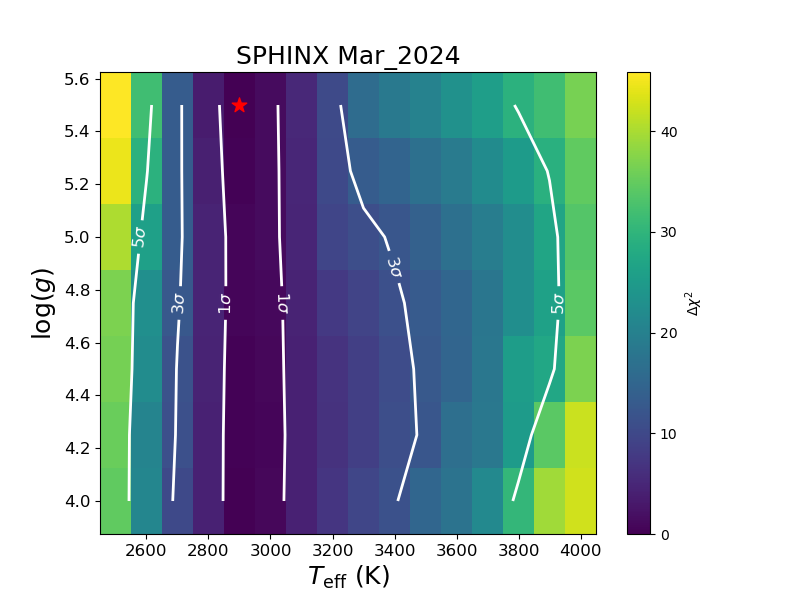}
    \includegraphics[width=0.47\textwidth,trim={0.7cm 0.3cm 1.4cm 0.6cm},clip]{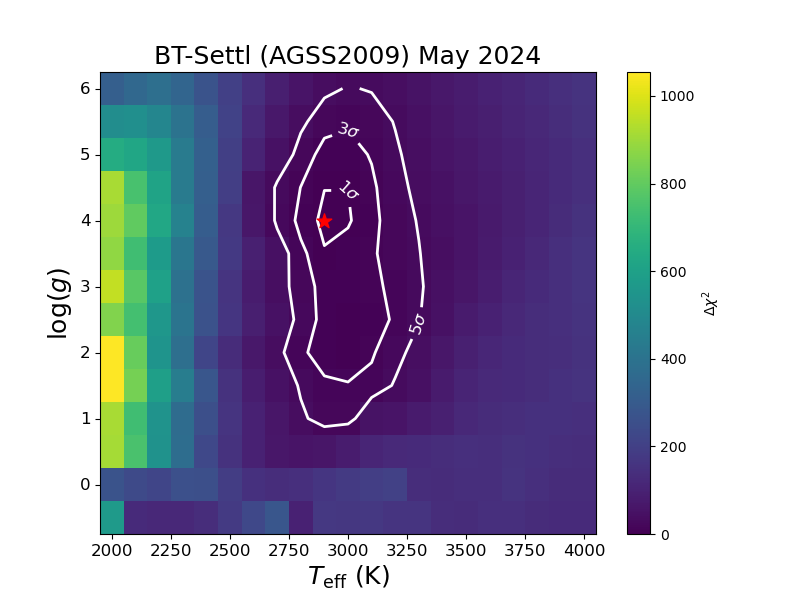}
    \includegraphics[width=0.47\textwidth,trim={0.7cm 0.3cm 1.4cm 0.6cm},clip]{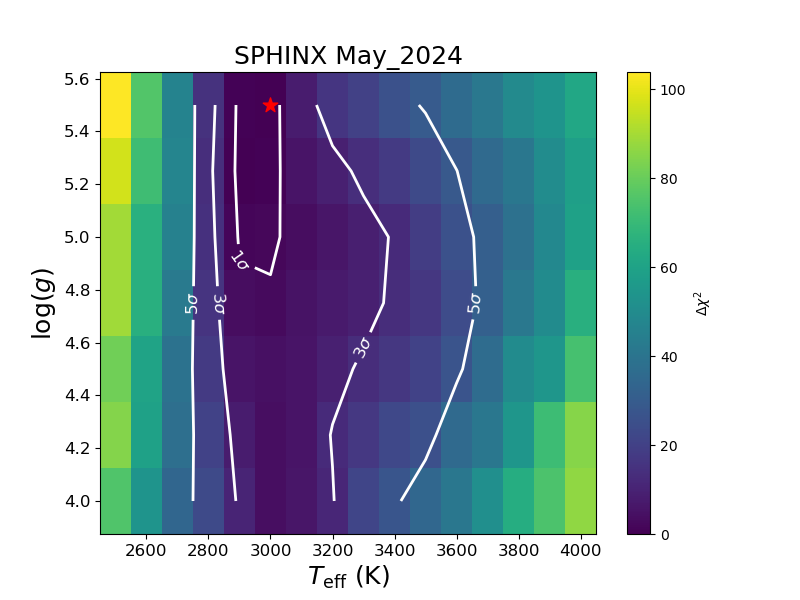}
    \caption{Contour plots of $\Delta\chi^2 = \chi^2 - \chi^2_{\min}$ containing the best-fit model points for BT-Settl (AGSS2009) (left) and SPHINX (right) for each epoch, with the 1, 3, and $5\sigma_{\Delta\chi^2}$ contours plotted. In each panel, the best-fit model to that epoch of observations (with $\chi^2 = \chi^2_{\min}$) is marked with a red star. Note that the large uncertainties on the CHARIS $JHK$ spectrum in the February 2022 epoch result in very small $\Delta\chi^2$ values for the best-fit models, resulting in particularly poor constraints on the log(g) parameter for that epoch that reflect the conservative uncertainties obtained. }
    \label{fig:Total_Contour_Fits}
\end{figure*}

The best-fit results of the models are shown in Table \ref{tab:model_fits} and visualized in Figure \ref{fig:Best_Model_Fits}, while Figure \ref{fig:Total_Contour_Fits} shows the contour plots as a function of $T_{eff}$ and $\log(g)$ including the best-fit model of each epoch of observation. Overlaid on the plots are contours signifying the $\left\lbrace1, 3, 5\right\rbrace \times\sigma_{\Delta\chi^{2}}$ values, where $\Delta\chi^{2} = \chi^2 - \chi^2_{\min}$ and $\chi^2_{\min}$ corresponds to the best-fit model spectrum for that specific model grid and epoch of observation. The $\Delta\chi^2$ thresholds that correspond to $\left\lbrace1, 3, 5\right\rbrace \times\sigma_{\Delta\chi^{2}}$ are calculated using a similar method to that used above in Section \ref{ss:spec-empirical}. In this case, when converting the Gaussian distribution to a $\chi^2$ distribution, we instead use a $\chi^2$ distribution with DoF$ = 2$. For the purposes of our analysis, the effective temperature $T_{eff}$ and gravity $\log(g)$ are considered relevant parameters following the procedure of \citep{1976ApJ...210..642A}. Although metallicity, $\alpha$-element abundance, and C/O values were parameters varied for the model fitting, the low resolution spectra was not designed to meaningfully constrain them, so they are not included in our degrees of freedom. As a result, we collapse each grid of model fits to a single, 2-dimensional grid over $T_{eff}$ and $\log(g)$ by only retaining the best-fit model for each $(T_{eff},\log(g))$ across all other parameters used in that grid. 
For the resulting DoF$=2$, $\left\lbrace 1, 3, 5 \right\rbrace \times \sigma_{\Delta\chi^2}$ correspond to $\Delta \chi^2 = \left\lbrace 2.30,  11.83, 28.74 \right\rbrace$.

One point of consideration is the limits implemented by the models; as noted in the captions, many of the contours unfortunately do not fully close within the span of log(g) values covered by these models. 
Notably, as with the comparison to the empirical spectral library in Section \ref{ss:spec-empirical}, the larger uncertainties of the February 2022 CHARIS spectrum result in a best-fit model with $\chi^2_{\nu} < 1$ (reduced using the original $\textrm{DoF}=17$ for the individual model fits), and lead to generally poorer constraints on the derived model parameters. While we include the results from the February 2022 epoch for completion, we emphasize that its low $\chi^2_{\nu} < 1$  likely indicates an overestimation of uncertainties in the observed spectrum from that epoch, rather than a better fit to the models. This is consistent with the wider $\left\lbrace 1, 3, 5 \right\rbrace \times \sigma_{\Delta\chi^2}$ contours for this epoch seen in Figure \ref{fig:Total_Contour_Fits}, as increasing the uncertainty of the measured spectrum allows a wider range of models to be consistent with the observations at a given level of significance.

The effective temperatures of the best-fit models across all three epochs of observations and two model grids are in agreement with each other, showing that the object has an effective temperature of around 3000~K. As shown by the contours in Figure \ref{fig:Best_Model_Fits}, the range of temperatures is relatively well constrained, with temperatures ranging from 2800-3100~K within $1\sigma_{\chi^2_\nu}$ for the 2024 epochs.

The $\log(g)$ values of the best-fit models vary significantly between the SPHINX and BT-Settl models. The BT-Settl models prefer slightly lower $\log(g)$, with the two 2024 epochs yielding a best fit to $\log(g) = 4.0$ and $1\sigma_{\chi^2_\nu}$ contours consistent with $\log(g)$ between $\sim 3.5$ and $4.5$. The SPHINX models favor higher $\log(g)$ values, with both 2024 epochs best fitting $\log(g) = 5.5$. However, The SPHINX models span a much smaller range of $\log(g)$ values than the BT-Settl models; as a result, they do not provide strong constraints on the $\log(g)$. For all epochs of our HIP~46065~B spectral observations, the best-fit value falls at the upper limit of the SPHINX model grid's coverage in $\log(g)$, precluding derivation of any upper limits on $\log(g)$ uncertainties. Only the May 2024 epoch provides constraints on the lower $1\sigma_{\chi^2_\nu}$ uncertainty for $\log(g)$, at $\sim 4.9$.

With our fitted values for $\alpha_f$ for Equation \eqref{eq:chisq} for each of the models, the radius of our best-fits for HIP 46065~B can be calculated using the parallax measurements from the Gaia DR3, as shown in Table \ref{tab:model_fits}. The model grid fits imply a companion radius of roughly 2 $R_{J}$.

We briefly compare the results of the SPHINX model fits to the Sonora Red Diamondback evolutionary models \citep{Davis2025}. The Red Diamondback models combine the SPHINX model atmospheres for M-dwarfs \citep{Iyer_2023} with the Sonora Diamondback brown dwarf models \citep{Morley2024} to derive a model for the early evolution of planetary, substellar, and low-mass stellar objects from $\sim0.5 \textrm{ M}_J$ to $\sim84 \textrm{ M}_J$. To accomodate this large range in object masses, the Red Diamondback models also include treatment for the onset of Deuterium and Hydrogen burning \citep{Davis2025}. Due to the combination of different spectral model grids, these evolutionary models only cover the SPHINX models with $C/O = 0.5$ and $[M/H]$ between $-0.5$ and $+0.5$. 

Assuming that the variation of age and mass as a function of $C/O$ and $[M/H]$ is small compared to the dependence on $T_{eff}$ and $\log(g)$, we use a multi-variate linear interpolator to derive a function for the mass and age as a function of $T_{eff}$ and $\log (g)$ for the Red Diamondback models. Using that interpolated function, we generate grids of Red Diamondback ages and masses across the same ($T_{eff}$, $\log(g)$) as the SPHINX spectral fits. Finally, we overlay the $\chi^2$ contour from the SPHINX spectral fit to the May 2024 data, to determine what these spectral constraints imply about the mass and age of the companion. The best-fit spectral model would correspond to a Red Diamondback age of $\sim 140$~Myr, with a mass of $\sim 82 \textrm{ M}_J$. While the lack of a constraint on the maximum $\log(g)$ at the $1\sigma$ level precludes a full constraint on the Red Diamondback age and mass, the constraints that fall within the ($T_{eff}$, $\log(g)$) range of the SPHINX grids correspond to a $68\%$ confidence interval that the age is above $70$~Myr and the mass is above $\sim 81\textrm{ M}_J$. However, we caution that, with the best-fit model being at the boundary of the available SPHINX grid, the actual best-fit model could lie beyond the bounds of the availabale SPHINX parameter space. As such, the Red Diamondback age and mass derived here should be considered only as approximate lower constraints, to check for consistency with stronger constraints derived from other methods. 
Additional spectral coverage, higher spectral resolution, or higher signal-to-noise would be necessary to further constrain the evolutionary mass of the companion derived from its spectrum.

\begin{deluxetable*}{llrrrr}[ht]
     \tablewidth{0pt}
    \tablecaption{HIP 46065 B Relative Astrometry\label{tbl:astrometry}}
    \tablehead{\colhead{UT Date} & \colhead{Instrument} &  \colhead{$\rho$ (mas)} &{$\sigma_{\rho}$ (mas)} & \colhead{$PA$ ($\degr$)} 
    & \colhead{$\sigma_{PA}$ ($\degr$)} }
    \startdata
    2022-02-22 & SCExAO/CHARIS & 660.7 &  3.4 & 124.38 & 0.38 \\
    2023-11-19 & Keck/NIRC2    & 661.5 & 12.3 & 124.65 & 0.32 \\
    2024-03-19 & SCExAO/CHARIS & 676.1 &  3.4 & 124.97 & 0.38 \\
    2024-05-02 & SCExAO/CHARIS & 675.5 &  3.5 & 125.26 & 0.38 \\
    2024-11-20 & SCExAO/CHARIS & 677.7 &  3.8 & 125.27 & 0.38 \\
    2024-12-17 & Keck/NIRC2    & 671.6 &  6.5 & 125.44 & 0.55 \\
    \enddata
    
    \end{deluxetable*}

\begin{deluxetable*}{lccc}
\tablecaption{MCMC Orbital Fit Results for HIP~46065 \label{tab:hip46065_orbit}}
\tablehead{
\colhead{Parameter} & \colhead{Median and 68\% CI} & \colhead{95\% CI} & \colhead{Prior}
}
\startdata
\multicolumn{4}{c}{\textit{Fitted Parameters}} \\
\hline
$M_{\rm pri}$ ($M_\odot$) & ${1.87}^{+0.30}_{-0.30}$ & $(1.252,\;2.475)$ & Gaussian \\
$M_{\rm sec}$ ($M_{\rm Jup}$) & ${47}^{+75}_{-25}$ & $(0.007,\;464.628)$ & uniform \\
Semimajor axis $a$ (au) & ${50}^{+23}_{-11}$ & $(34.605,\;108.361)$ & uniform \\
$\sqrt{e}\sin\omega$ & ${0.02}^{+0.51}_{-0.53}$ & $(-0.855,\;0.876)$ & uniform \\
$\sqrt{e}\cos\omega$ & ${0.34}^{+0.46}_{-0.93}$ & $(-0.877,\;0.94)$ & uniform \\
Inclination ($^\circ$) & ${64}^{+16}_{-29}$ & $(13.939,\;94.893)$ & $\sin i$ (geometric) \\
PA of ascending node $\Omega$ ($^\circ$) & ${301}^{+23}_{-174}$ & $(23.82,\;346.909)$ & uniform \\
Mean longitude ($^\circ$) & ${124}^{+170}_{-46}$ & $(25.527,\;332.511)$ & uniform \\
Parallax (mas) & ${11.050}^{+0.014}_{-0.014}$ & $(11.022,\;11.077)$ & Gaussian, Gaia DR3 \\
\hline
\multicolumn{4}{c}{\textit{Derived Parameters}} \\
\hline
Period (yrs) & ${261}^{+198}_{-85}$ & $(142.718,\;784.345)$ & -- \\
Argument of periastron $\omega$ ($^\circ$) & ${175}^{+156}_{-142}$ & $(4.791,\;355.592)$ & -- \\
Eccentricity $e$ & ${0.66}^{+0.23}_{-0.40}$ & $(0.048,\;0.977)$ & -- \\
Semimajor axis (mas) & ${557}^{+252}_{-124}$ & $(382.421,\;1197.487)$ & -- \\
Periastron time $T_0$ (JD) & ${2519733}^{+45116}_{-18383}$ & $(2486858.218,\;2662957.343)$ & -- \\
Mass ratio $q$ & ${0.025}^{+0.039}_{-0.014}$ & $(0.0,\;0.25)$ & -- \\
\enddata
\end{deluxetable*}

\section{Astrometry and Orbit Modeling\label{sec:orbit}}

As discussed in Section \ref{sec:intro}, orbital modeling of companions to accelerating stars provides dynamical mass constraints independent of evolutionary models, allowing the system to serve as a benchmark for model verification and calibration. The high-contrast observations of HIP~46065~B presented here span little relative orbital motion from the companion, due in combination to the larger distance from the host star and the short time elapsed since discovery. In combination with other dynamical measurements of its host star across a longer baseline, however, initial constraints on the independent, dynamical mass of the companion may be derived. 
We jointly fit the relative astrometry, Gaia DR3 proper motion anomaly, and Hipparcos-Gaia acceleration constraints using the \texttt{orvara} code \citep{Brandt2021}, which employs a parallel-tempering MCMC sampler to efficiently explore the multidimensional parameter space. The fitted parameters include the standard Keplerian orbital elements: semimajor axis ($a$), eccentricity ($e$), inclination ($i$), longitude of ascending node ($\Omega$), argument of periastron ($\omega$), and epoch of periastron passage ($T_0$). 

We adopt Gaussian priors on the stellar mass of HIP~46065~A of 
$M_A = 1.91 \pm 0.3~M_\odot$. While the peak of this distribution is directly from \citet{Kervella_2022}, we widen the $\sigma$ for the prior from the uncertainty quoted therein. For less-studied field stars like HIP~46065, photometric masses can depend significantly on the population-level assumptions, which may or may not be accurate for that particular source. Using these photometric masses as strongly-constrained, informative priors during orbital modeling can bias the resulting best-fit parameters, including dynamical masses. As the goal of orbit modeling is to derive independent dynamical masses, we opt for a less-informative prior on the primary mass by widening the gaussian distribution to $\sigma = 0.3~M_\odot$. The resulting distribution is consistent with a range of previously-published photometric masses (including $1.98 \pm 0.12~M_\odot$, \citealt{AllendePrieto1999}; $2.13 \pm 0.33~M_\odot$, \citealt{Stassun2019}; \centmass, Section \ref{sec:age}). Uniform priors are used for $\cos i$, $\omega$, and $\Omega$. 

The orbit fit combines the relative astrometry 
from SCExAO/CHARIS and Keck/NIRC2 (Table \ref{tbl:astrometry}) with the absolute acceleration of 
HIP~46065~A inferred from the Hipparcos--Gaia Catalog of Accelerations 
(HGCA; \citealt{Brandt2021b}). This joint treatment provides sensitivity 
to the full three-dimensional orbit despite the relatively short 
direct-imaging time baseline. We explore the parameter space with the 
parallel-tempered MCMC sampler implemented in \texttt{orvara}, running 
$5\times10^5$ total steps across 30 temperature chains to ensure robust 
mixing and convergence. Posterior stationarity is verified through the 
Gelman--Rubin statistic ($\hat{R} < 1.02$ for all parameters) together 
with visual inspection of the trace plots.

\begin{figure*}[ht]
    \centering
\includegraphics[width=0.35\textwidth]{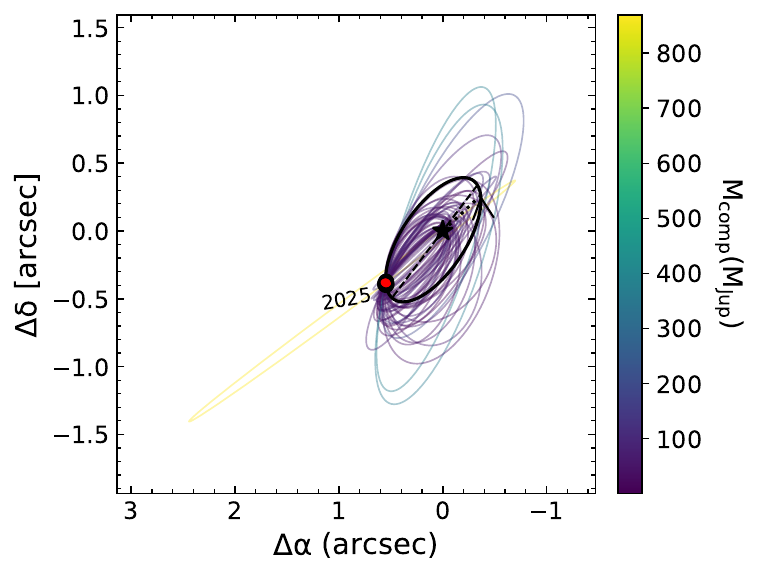}
\includegraphics[width=0.315\textwidth]{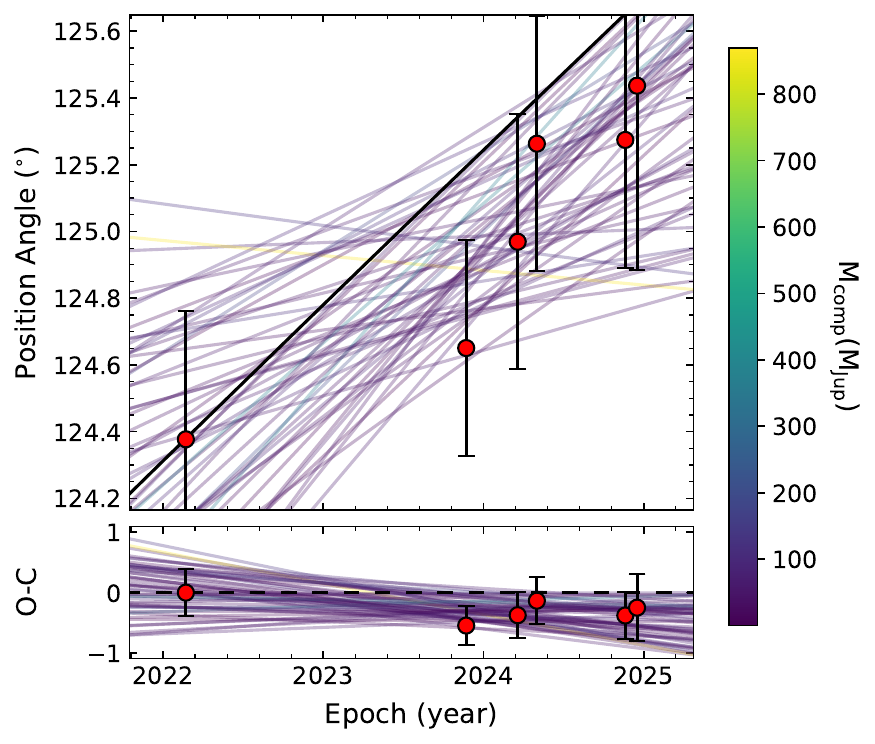}
\includegraphics[width=0.32\textwidth]{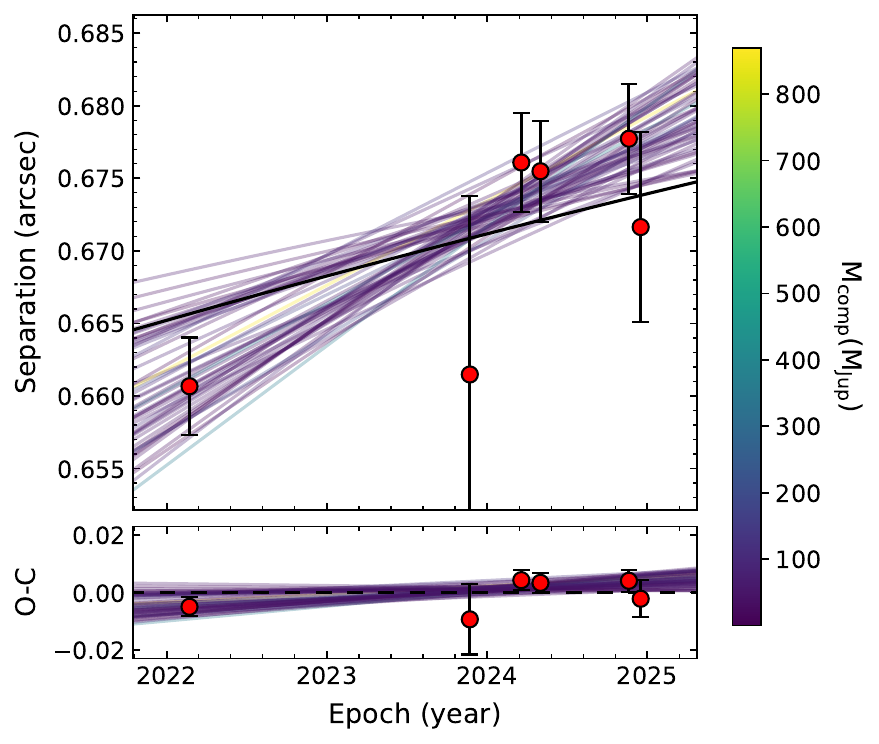}
    \caption{Astrometric residuals for HIP~46065~B relative to the best-fit orbit. 
    From left to right: sky-projected orbit, observed minus computed position angle, observed minus computed separation. 
    Residuals remain within the adopted measurement uncertainties (a few mas and tenths of a degree).}
    \label{fig:hip46065_OC}
\end{figure*}

\begin{figure*}[ht]
\centering
\includegraphics[width=\textwidth]{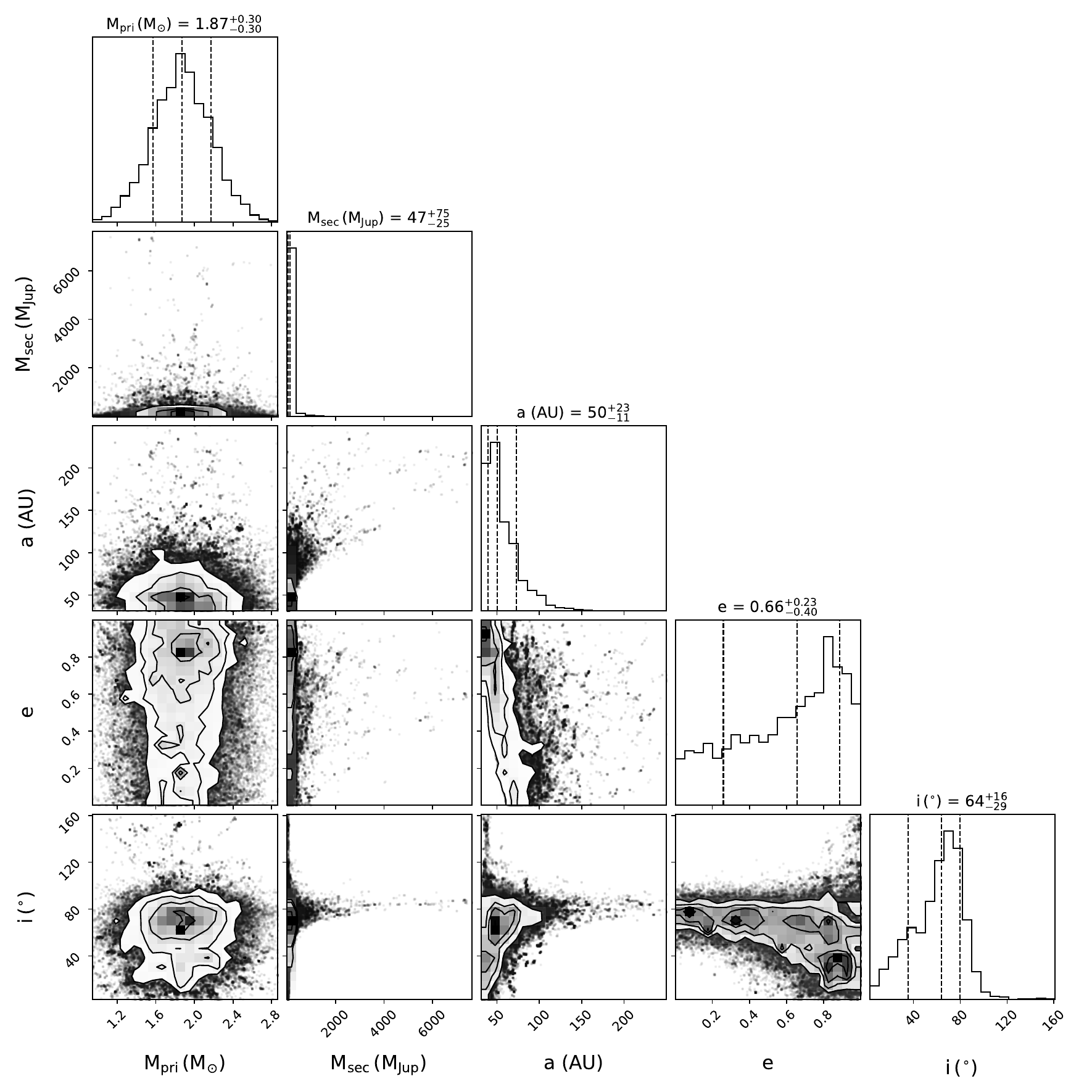}
\caption{Posterior distributions for orbital parameters of HIP~46065~B derived from the joint astrometric fit using \texttt{orvara}. Contours represent the 1$\sigma$, 2$\sigma$, and 3$\sigma$ confidence levels.}
\label{fig:hip46065_corner}
\end{figure*}

The combined fit yields a bound, moderately eccentric companion on a
well-constrained orbit. We find a semimajor axis of
$a = 50^{+23}_{-11}~\mathrm{AU}$ and an eccentricity of
$e = 0.66^{+0.23}_{-0.40}$, corresponding to an orbital period of
$P = 261^{+198}_{-85}~\mathrm{yr}$. The inclination of
$i = 64^{+16}_{-29}~\mathrm{deg}$ indicates that the orbit is moderately
tilted relative to the sky plane. The longitude of the ascending node and
argument of periastron are
$\Omega = 301^{+23}_{-174}~\mathrm{deg}$ and
$\omega = 175^{+156}_{-142}~\mathrm{deg}$, respectively. The dynamical
companion mass is
$M_B = 47^{+75}_{-25}~M_{\rm Jup}$, consistent with the spectral type and effective temperatures derived 
from atmospheric model fitting (Sections~\ref{ss:spec-empirical} and \ref{ss:spec-model}). The best-fit
orbit and full posterior distributions are shown in
Figures~\ref{fig:hip46065_OC}--\ref{fig:hip46065_corner}. The moderately inclined orbit of $i \approx 64^\circ$ suggests that the
system is neither close to face-on nor edge-on. 

While this provides initial constraints on the model-independent dynamical mass of the companion, the 95\% confidence interval on the companion mass spans $(0.007, 465)~M_{\textrm{Jup}}$, as shown in Table \ref{tbl:astrometry}. The large
uncertainties on the companion mass and other derived parameters -- and particularly on $\Omega$ and $\omega$ -- primarily reflect the limited coverage of
the orbital phase in current imaging data available since the companion's discovery in our 2022 SCExAO/CHARIS observations. Observations spanning a longer time baseline will significantly refine these constraints, particularly in the form of relative astrometry from high-contrast imaging and companion radial velocity measurements.

\section{The Age of the HIP~46065 System\label{sec:age}}

To determine the mass of the companion with luminosity through evolutionary models, it is necessary to estimate the age of the system.

\subsection{Metallicity Determination for HIP 46065 A \label{subsec:hostmetal}}

Metallicity of the host star can be a powerful prior when using evolutionary models to determine age. We utilize the approach from \cite{Baburaj2026}, which uses the PHOENIX atmosphere models \citep{Husser2013} and high-resolution ($R\sim31,500$) APO/ARCES \citep{ARCES_2003} spectra for the primary HIP 46065 A to estimate its metallicity. A MCMC sampling approach using \texttt{emcee} \citep{foreman-mackey_emcee_2013}, as incorporated in the Python module \texttt{SMART} \citep{Hsu2021} is used to fit the PHOENIX models to our spectra. Spectral orders corresponding to wavelengths 5850--5950 \r{A} (containing the Na \textsc{I} D doublet), 5600--5700 \r{A}, and 6100--6200 \r{A} (the latter two containing several iron/metal lines) are used to obtain [M/H]$\,=-0.02^{+0.10}_{-0.06}$ for HIP 46065 A.

\subsection{Bayesian Age Determination for HIP 46065} 

Subsequently, we infer HIP~46065's age using the software, \texttt{BELFAST} (a Bayesian framework for Estimating the ages of earLy-type stars with FAST rotation; Gu et al. in preparation). 
We use \texttt{BELFAST} to model the photometric measurements and projected rotation velocity ($v\sin{i}$) of HIP~46065 with the PARSEC V2.0 evolutionary tracks and isochrones, which account for rotation in evolutions \citep{nguyen_parsec_2022, Nguyen_2025_PARSEC}.
The Bayesian inference progresses via MCMC sampling performed by \texttt{emcee} \citep{foreman-mackey_emcee_2013}.  
We included the photometric measurements of $G= 6.7175 \pm 0.0002$ mag, $G_{\rm{BP}}= 6.7551 \pm 0.0005$ mag, and $G_{\rm{RP}} = 6.6200 \pm 0.0004$ mag from Gaia DR3 \citep{GAIA2023}; $B_{\rm{T}}= 6.869 \pm 0.014$ mag and $V_{\rm{T}}= 6.737 \pm 0.010$ mag from Tycho2 \citep{Hog_2000_tycho2}; and $J = 6.503 \pm 0.024$ mag, $H = 6.511 \pm 0.033$ mag, and $K_{\rm{s}} = 6.478 \pm 0.024$ mag from 2MASS \citep{skrutskie_two_2006}.
In addition, we included a line broadening parameter $v_{\rm broad} = 180.22 \pm 4.53 \rm \: km\,s^{-1}$ from Gaia DR3 in our fit with an assumption that $v_{\rm broad}$ is equal to $v\sin{i}$. 
The fit also samples a parallax with a Gaussian prior that is based on a Gaia-DR3 parallax ($\varpi = 11.0497 \pm 0.0222$ mas). 
The age was sampled using a uniform prior between 10 and 2001 Myr. 
Meanwhile, we corrected an interstellar extinction sampled with a Gaussian prior with the mean and standard deviation derived from the \texttt{G-tomo} tool \citep{vergely2022three, lallement2022updated}, which yielded $A(550\,\mathrm{nm}) = 0.028 \pm 0.001\,\mathrm{mag}$, assuming $A(550\,\mathrm{nm}) = A_{V\,\mathrm{band}}$.
The interstellar extinctions over $B_{\rm{T}}$-to-$K_{\rm{s}}$ wavelengths were corrected using the extinction law that is based on \citet{wang_optical_2019}, which was slightly updated to apply it to the  $B_{\rm{T}}$ and $V_{\rm{T}}$ bands (see Gu et al. for detail). 
We adopted the same priors as adopted in \citet{brandt_bayesian_2015} except for the metallicity and the aforementioned parameters.  
Our metallicity prior is a Gaussian distribution with a mean of $\rm [M/H]=-$0.08 dex and a standard deviation of 0.10 dex,
which is based on our metallicity determination with the PHOENIX atmosphere models (see Section \ref{subsec:hostmetal}). 
We then conservatively rounded the uncertainties from the original measurements. 
While the PARSEC v2.0 models adopts the metallicity zero point from \citet{Caffau_2011_SolarAbundance}, the PHOENIX atmospehre adopts that from \citet{Asplund_2009}.
The metallicity prior adopted in this work accounts for this difference and the uncertainties of the zero points\footnote{We re-derived the solar metallicity $Z_{\rm{\odot}}$ adopted in the PHOENIX model using the information in \citet{Husser2013} and the python module \texttt{periodictable} \citep{kienzle_2026_periodictable}.}.

\begin{figure}
    \centering
    \includegraphics[width=\columnwidth]{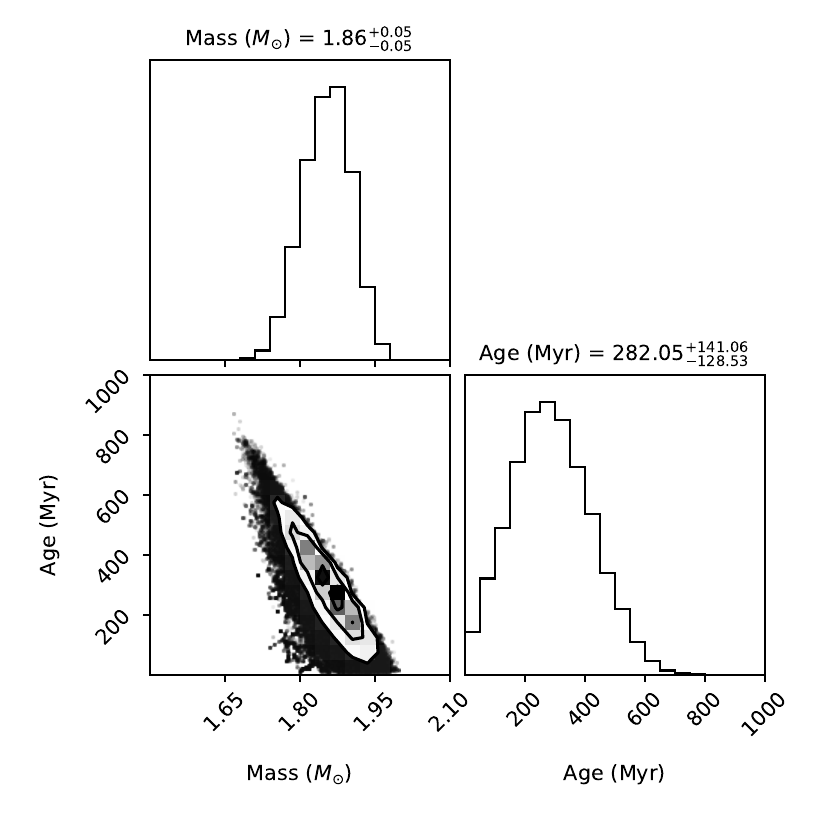}
    \caption{Mass and age posterior map obtained from our Bayesian age inference for HIP 46065.}
    \label{fig:Mass_Age_HIP46065}
\end{figure}

We use \texttt{BELFAST} (Gu et al. in preparation) to calculate brightness magnitudes dependent on a rotation-axis inclination at each MCMC step. 
The PARSEC v2.0 models include gravity darkening, and provide inclination-dependent photometric grids. 
We ran \texttt{emcee} with 40 walkers, using a combination of 95\% StretchMove and 5\% KDEMove moves to improve the sampling efficiency. 
We discarded the initial 73240 samples as burn-in samples and evaluated the convergence of MCMC sampling by computing the integrated autocorrelation time ($\tau_{\rm{f}}$); the total number of samples exceeds $70\tau_{\rm{f}}$. 
As a result, we obtained the mass-age posterior map plotted in Figure \ref{fig:Mass_Age_HIP46065} and estimated the mass and age of HIP 46065 to be \centmass and \centage, respectively.
We also found that 95\% of samples fall in the age range of 50--520 Myr.

\section{Summary\label{sec:disc}}

We report the discovery of a low mass, candidate substellar companion to the astrometrically-accelerating star, HIP~46065. A comparison of its measured $JHK$ spectrum to sources in the SpeX Prism library suggest a NIR spectral type of M5$_{-1}^{+3}$, with known subdwarfs and extreme subdwarfs excluded at the $> 3 \sigma $ level. Comparison with the BT-Settl (AGSS2009) stellar models \citep{2011ASPC..448...91A} and the SPHINX M-dwarf spectral models \citep{Iyer_2023} support this conclusion, indicating the companion has an effective temperature of $3000^{+100}_{-200}$~K. The BT-Settl models indicate a $\log(g) = 4.0 \pm 0.5$. The SPHINX models prefer higher $\log(g) \sim 5.5$ to best fit our HIP~46065~B observations; however, our observations do not have sufficient spectral resolution to further constrain the SPHINX-indicated $\log(g)$ within the range of model values provided by the SPHINX grid.

Joint fitting of nearly 3 years of relative astrometry and the Hipparcos-Gaia acceleration constraints of HIP~46065~A using \texttt{orvara} yields a dynamical mass of $M_B = 47 _{-25}^{+75} M_{Jup}$, with uncertainties straddling the stellar/substellar boundary. This is consistent with the $M_B \gtrsim 81 M_J$ mass constraint derived from a comparison of the SPHINX spectral analysis to the Red Diamondback evolutionary models. Additional orbital coverage is required to further constrain its mass with respect to the H-burning limit. 

We also constrain the age of the host star, HIP~46065~A, using the upcoming Bayesian inference software for early-type stars, \texttt{BELFAST} (Gu et al., in preparation). We derive an age of  \centage\ and a mass of  \centmass\, for the host star. 
This is consistent with the dynamical primary star mass derived from the orbital modeling in Section \ref{sec:orbit} ($M_{A,orbit} = 1.87 \pm 0.30 M_{\odot}$). 
To this end, we also obtain high-resolution ($R \sim 31,500$) APO/ARCES \citep{ARCES_2003} spectra for the primary, HIP~46065~A, and derive its metallicity by using MCMC sampling to fit the PHOENIX stellar models \citep{Husser2013} to the observed stellar spectrum. We derive a metallicity of $\textrm{[M/H]}=-0.02^{+0.10}_{-0.06}$ for HIP~46065~A.

\begin{acknowledgements}
\indent We thank the anonymous referee for thorough and helpful comments and suggestions that improved the quality of this paper.

\indent The authors acknowledge the very significant cultural role and reverence that the summit of Mauna Kea holds within the Hawaiian community.  We are most fortunate to have the opportunity to conduct observations from this mountain.   

\indent The development of SCExAO was supported by JSPS (Grant-in-Aid for Research \#23340051, \#26220704 \& \#23103002), Astrobiology Center of NINS, Japan, the Mt Cuba Foundation, and the director's contingency fund at Subaru Telescope.  CHARIS was developed under the support by the Grant-in-Aid for Scientific Research on Innovative Areas \#2302. The development of CACAO, which runs the adaptive optics control loop, is supported by the National Science Foundation under award 2410616. Some of the data presented herein were obtained at the W. M. Keck Observatory, which is operated as a scientific partnership among the California Institute of Technology, the University of California and the National Aeronautics and Space Administration. The Observatory was made possible by the generous financial support of the W. M. Keck Foundation. M.K. is supported by JSPS KAKENHI (grant No. 24K07108). Z.G. is supported by FoPM, WINGS Program, the University of Tokyo. N.S. acknowledges support from the Heising-Simons Foundation. \\

\software{
    \texttt{ADEPTS} \citep{Tobin2020, Tobin2022},
    \texttt{Astropy} \citep{astropy1, astropy2},
    \texttt{Astropy Specutils FluxConservingResampler} \citep{specutils,Carnall2017},
    \texttt{BELFAST} (Gu et al. in preparation),
    CHARIS Data Processing Pipeline (\texttt{DPP}; \citealt{Currie2020b}),
    CHARIS Data Reduction Pipeline (\texttt{DRP}; \citealt{Brandt2017}),
    \texttt{emcee} \citep{foreman-mackey_emcee_2013},
    \texttt{NumPy} \citep{numpy},
    \texttt{orvara} \citep{Brandt2021},
    \texttt{SciPy} \citep{scipy},
    \texttt{SMART} \citep{Hsu2021, SMART},
    SpeX Prism Library Analysis Toolkit (\texttt{SPLAT}; \citealt{SPLAT}),
    \texttt{Matplotlib} \citep{matplotlib,matplotlib3.4.3},
    \texttt{periodictable} \citep{kienzle_2026_periodictable},
    \texttt{corner} \citep{Foreman-Mackey_2016_corner}
    }

\end{acknowledgements}

\bibliographystyle{aasjournalv7}
\bibliography{bibliography}

\end{document}